\documentclass[useAMS,usenatbib]{mnras}
\usepackage{graphicx}
\usepackage{times}
\usepackage{amssymb}
\usepackage{amsmath}
\usepackage{multirow}
\usepackage{subfig}
\usepackage{hyphenat}
\usepackage{sansmath}
\usepackage{aas_macros}
\usepackage{tabularx}
\usepackage{widetext}
\newcommand{\hompc}{\,h\,{\rm Mpc}^{-1}}
\newcommand{\mpcoh}{\,h^{-1}\,{\rm Mpc}}

\newcommand{\bk}{\boldsymbol{k}}
\newcommand{\bq}{\boldsymbol{q}}

\newcommand{\br}{\boldsymbol{r}}

\newcommand{\bC}{\boldsymbol{\mathsf{C}}}

\makeatletter
\let\origsection\section
\let\origsubsection\subsection
\let\origsubsubsection\subsubsection
\renewcommand\section{\@ifstar{\starsection}{\nostarsection}}
\renewcommand\subsection{\@ifstar{\starsubsection}{\nostarsubsection}}
\renewcommand\subsubsection{\@ifstar{\starsubsubsection}{\nostarsubsubsection}}
\newcommand\nostarsection[1]{\sectionprelude\origsection{#1}\sectionpostlude}
\newcommand\nostarsubsection[1]{\subsectionprelude\origsubsection{#1}\subsectionpostlude}
\newcommand\nostarsubsubsection[1]{\subsubsectionprelude\origsubsubsection{#1}\subsubsectionpostlude}
\newcommand\starsection[1]{\sectionprelude\origsection*{#1}\sectionpostlude}
\newcommand\starsubsection[1]{\subsectionprelude\origsubsection*{#1}\subsectionpostlude}
\newcommand\starsubsubsection[1]{\subsubsectionprelude\origsubsubsection*{#1}\subsubsectionpostlude}
\newcommand\sectionprelude{\vspace{-1em}}
\newcommand\sectionpostlude{\vspace{0mm}}
\newcommand\subsectionprelude{\vspace{-1em}}
\newcommand\subsectionpostlude{\vspace{0mm}}
\newcommand\subsubsectionprelude{\vspace{-1em}}
\newcommand\subsubsectionpostlude{\vspace{0mm}}
\makeatother

\begin{document}

\title[Covariance Matrix Estimation] 
{Galaxy 2-Point Covariance Matrix Estimation for Next Generation Surveys}

\author[C. Howlett \& W. J. Percival]{\parbox{\textwidth}{
Cullan Howlett\thanks{Email: cullan.howlett@icrar.org}$^{1,2,3}$ \&
Will J. Percival$^{3}$
}
  \vspace*{4pt} \\ 
$^{1}$International Centre for Radio Astronomy Research, The University of Western Australia, Crawley, WA 6009, Australia. \\
$^{2}$ARC Centre of Excellence for All-sky Astrophysics (CAASTRO). \\
$^{3}$Institute of Cosmology \& Gravitation, Dennis Sciama Building, University of Portsmouth, Portsmouth, PO1 3FX, UK. \\
}

\pagerange{\pageref{firstpage}--\pageref{lastpage}} \pubyear{2017}
\maketitle
\label{firstpage}

\begin{abstract}
We perform a detailed analysis of the covariance matrix of the spherically averaged galaxy power spectrum and present a new, practical method for estimating this within an arbitrary survey \textit{without} the need for running mock galaxy simulations that cover the full survey volume. The method uses theoretical arguments to modify the covariance matrix measured from a set of small-volume cubic galaxy simulations, which are computationally cheap to produce compared to larger simulations and match the measured small-scale galaxy clustering more accurately than is possible using theoretical modelling. We include prescriptions to analytically account for the window function of the survey, which convolves the measured covariance matrix in a non-trivial way. We also present a new method to include the effects of supersample covariance and modes outside the small simulation volume which requires no additional simulations and still allows us to scale the covariance matrix. As validation, we compare the covariance matrix estimated using our new method to that from a brute force calculation using 500 simulations originally created for analysis of the Sloan Digital Sky Survey Main Galaxy Sample (SDSS-MGS). We find excellent agreement on all scales of interest for large scale structure analysis, including those dominated by the effects of the survey window, and on scales where theoretical models of the clustering normally break-down, but the new method produces a covariance matrix with significantly better signal-to-noise. Although only formally correct in real-space, we also discuss how our method can be extended to incorporate the effects of Redshift Space Distortions.
\end{abstract}

\begin{keywords}
cosmology: theory - large-scale structure of Universe
\end{keywords}

\section{Introduction}
For random-phase, Gaussian distributed density perturbations, all the cosmological information is included in the 2-point functions. Although gravitational evolution (and, if it exists, primordial non-Gaussianity) introduces phase-space information and small higher-order n-point functions, the majority of available information is still encapsulated in just the 2-point functions. The former of these can be readily measured using large surveys of the universe. However, the covariance matrix, which quantifies the error on the universe's power spectrum or correlation function, cannot be measured so easily.

As a result, the need to model the covariance matrix has become one of the most computationally demanding aspects of modern large scale structure analysis. Although this can be calculated analytically in the linear regime \citep{Feldman1994,Tegmark1997}, the non-linear galaxy covariance matrix is a complex function of non-linear shot-noise, galaxy evolution and the unknown relationship between the galaxies and the underlying dark matter. In any real application this is further complicated by the effect of Redshift Space Distortions (RSD). Recent progress has been made in understanding and computing the dark-matter covariance matrix theoretically \citep{Neyrinck2011,Mohammed2014,Carron2015,Bertolini2016,Mohammed2017,Barreira2017a,Barreira2017b}, but large simulation suites show that much work still needs to be done to understand the small-scale evolutionary effects \citep{Takahashi2009,Li.Y2014a,Blot2015,Klypin2017}, let alone modelling the \textit{galaxy} covariance matrix we actually measure. A more common solution is to use a set of detailed galaxy simulations, otherwise known as mock catalogues (mocks), to calculate a brute-force estimate of the covariance matrix.

In recent large scale structure analyses this estimation was performed using large numbers of simulations that cover the full survey volume, in both the angular and radial directions, with high enough resolution to accurately reproduce the galaxies within the survey. Earlier work, such as that of the 2dF Galaxy Redshift Survey (2dFGRS; \citealt{Colless2001, Colless2003}) used Log-normal realisations of the overdensity field (LN; \citealt{Coles1991, Cole2005}). The more recent SDSS-III Baryon Oscillation Spectroscopic Survey (BOSS; \protect{\mbox{\citealt{Anderson2012,Anderson2014, Alam2016}}}) used more sophisticated methods such as PTHALOS \protect{\mbox{\citep{Scoccimarro2002, Manera2013, Manera2015}}}, Quick Particle Mesh Simulations \protect{\mbox{(QPM; \citealt{White2014})}} and Augmented Lagrangian Perturbation Theory \protect{\mbox{(ALPT; \citealt{Kitaura2013})}} to produce their mock catalogues. Other alternatives include PINOCCHIO \protect{\mbox{\citep{Monaco2013,Munari2016}}}, Effective Zel’dovich approximation mocks (EZmocks; \citealt{Chuang2015a}), the Comoving Lagrangian Acceleration method (COLA; \citealt{Tassev2013, Tassev2015, Howlett2015b}) and the work of \cite{Sunayama2016}. \cite{Chuang2015b} and \cite{Monaco2016} provide reviews of the above methods, detailing their respective strengths and weaknesses, whilst work to create ever faster algorithms continues. Ultimately, this abundance of different methods attests to the increasing urge to reduce the computational burden of covariance matrix estimation. 

However, for now this burden will only be exacerbated by future surveys. Reaching the desired non-linear accuracy for the Dark Energy Spectroscopic Instrument (DESI; \citealt{Levi2013}), the Large Sky Synoptic Telescope (LSST; \citealt{Ivezic2008}) and Euclid \citep{Laureijs2011} may require more complex computational methods than are currently used. The covariance matrix also depends on cosmology; either an ensemble of simulations must be run for each model of interest, the covariance matrix from a single cosmology must be interpolated for other models \citep{White2015}, or the likelihood distribution from the comparison between the model and data must be modified \citep{Kalus2016}. In addition to this, the number of simulations in an ensemble may also need to increase. Recent studies by \cite{Dodelson2013, Taylor2013} and \cite{Percival2014} have shown that $\mathcal{O}(1000)$ mocks are required to obtain an accurate numerical estimate of the covariance matrix with sub-dominant errors compared to the statistical errors themselves for current surveys. However as the statistical errors in measurements of the galaxy clustering decrease, the number of simulations must increase to ensure the precision on the covariance matrix remains subdominant.

This presents a bleak picture for the standard method of covariance matrix estimation, in which a delicate balance between the speed, size and accuracy of each simulation must be achieved. Using the brute force approach, enough simulations must be run to estimate the covariance matrix to high precision, but they must also be large enough to fit the survey and have enough particles to reproduce the galaxy population. There have been many studies recently aiming to ease this problem by reducing the amount of simulations required to reach a given covariance matrix precision, rather than simply increasing the speed with which each realisation of the survey can be produced.

For a fixed simulation size, one technique for reducing the number of simulations required to reach a given covariance matrix precision is covariance matrix tapering \citep{Paz2015} where the covariance matrix is made more diagonal through the use of a specialised set of tapering functions. \cite{Padmanabhan2016} also present a method to directly estimate the inverse covariance matrix from simulations, which improves convergence in the estimate with the number of simulations. Both of these use the fact that the covariance matrix is generally sparse and contains off-diagonal terms that have low signal-to-noise. Other methods \citep{Schafer2005, Pope2008, OConnell2016, Pearson2016} combine an empirical estimate of the covariance matrix from a small number of samples with fitting functions containing several free parameters, whilst \cite{Cole1997} and \cite{Schneider2011} presented a method to add large-scale modes to small-scale simulations, thus enabling the fast creation of many full-size approximate simulations. All these methods succeed in greatly reducing the number of mock catalogues required to reach a given covariance matrix precision, however often contain free parameters which must be calibrated. Furthermore, these methods do not overcome the problem that running even a few hundred simulations may be a challenge for next generation surveys.

Instead of reducing the number of mocks required to obtain the covariance matrix to some accuracy, we propose a method to reduce the size of the simulations required to estimate the covariance matrix, utilising the known analytic properties of the covariance matrix, namely it's scaling with the volume of the simulation. Such a method has been suggested recently by \cite{Escoffier2016} (although this paper does not explicitly refer to any volume scaling), \cite{Mohammed2017} and \cite{Klypin2017}, but here we provide a viable algorithm to do so. Our method also includes the effects of the survey window function, which cannot be naively included in the same way as when we have simulations that fit the full survey volume, and the effects of modes missing from small volume simulations that occur naturally in larger volumes. Compared to \cite{Schneider2011}, our approach should be more robust as we alter the parameters of the small-volume simulations rather than trying to adjust the results of simulations run with fixed parameters. We also analytically add the large-scale modes rather than doing this numerically, avoiding the need to simultaneously model the largest and smallest scales, and the resulting degradation of resolution. The overall benefit of our method is that we can use it in addition to methods detailed above to reduce the necessary number of mocks, and as we will show, we can even improve the accuracy of the estimated covariance matrix at fixed computational cost by running larger numbers of smaller simulations. 

The layout of this work is as follows: In Section~\ref{sec:motivation}, we outline our approach and demonstrate that reducing the volume of the simulations used to estimate the covariance matrix can reduce the computational time required to achieve a given precision in the estimation, or that conversely we can improve our estimate given a fixed computational time. In Sections~\ref{sec:supersamp} and \ref{sec:window}, we present methods to account for the lack of large scale modes and the survey window function. Finally we tie everything together in Section~\ref{sec:endresult}, demonstrating that our method using small volume simulations can recover the same covariance matrix as a brute force estimation using full-size simulations.

\section{Motivation: The covariance matrix and its error} \label{sec:motivation}

If the density perturbations present in the universe are drawn from a Gaussian distribution then the estimated power spectrum must be drawn from a chi-squared distribution and the estimated covariance matrix $\hat{\bC}$ from its higher-dimensional counterpart, the Wishart distribution,
\begin{equation}
\mathcal{P}(\hat{\bC}|\bC,n,p) = \left(\frac{|\hat{\bC}|^{\frac{n-p-1}{2}}}{2^{\frac{np}{2}}|\bC|^{\frac{n}{2}}\Gamma_{p}\left(\frac{n}{2}\right)}\right)e^{-\frac{1}{2}\mathrm{Tr}[\hat{\bC}\bC^{-1}]},
\end{equation}
where $\mathcal{P}$ gives the probability of measuring a $p\times p$ covariance matrix based on the true underlying covariance matrix $\bC$. $n$ is the number of degrees of freedom, which in the case of covariance matrices estimated from a set of mocks is $n=N_{s}-p-1$, where $N_{s}$ is the number of simulations and $p$ is the number of measurement bins. $\Gamma_{p}$ is the multivariate gamma function.

The covariance of the Wishart distribution is given by
\begin{equation}
\langle \Delta\hat{\mathsf{C}}_{i,j}\Delta\hat{\mathsf{C}}_{k,m} \rangle = n^{-1}(\mathsf{C}_{i,k}\mathsf{C}_{j,m}+\mathsf{C}_{i,m}\mathsf{C}_{j,k}).
\end{equation}
In the simplified case of a Gaussian random field where the covariance matrix is diagonal, this reduces to
\begin{equation}
\Delta\hat{\mathsf{C}}_{i,i} = \sqrt{\frac{2}{n}}\mathsf{C}_{i,i}.
\label{eq:coverr}
\end{equation}
Hence the error on the covariance matrix scales as one over the square root of the number of degrees of freedom. This scaling has been tested and verified even for non-linear simulations by \cite{Takahashi2009} and \cite{Takahashi2011}. For a number of mocks much larger than the number of measurement bins, the precision of the covariance matrix is doubled if four times more mocks are used. Overall, if the covariance matrix is estimated only from simulations, the number of mocks required to reach the necessary covariance matrix precision for next generation surveys will be much larger than the number currently used.

However, the error on the covariance matrix depends on covariance matrix itself. It has also long been established that the covariance matrix scales as the inverse of the volume in which the power spectrum is measured \citep{Feldman1994, Meiksin1999, Scoccimarro1999}, where in the absence of a window function
\begin{align}
\mathsf{C}^{sm}(k_{i},k_{j}) & = \frac{2(2\pi)^{3}}{V_{k_{i}}V}\biggl(\bar{P}(k_{i})+\frac{1}{\bar{n}}\biggl)^{2}\delta^{D}(k_{i}-k_{j}) \notag \\
& + \frac{2}{\bar{n}^{2}V}\biggl(\bar{P}(k_{i})+\bar{P}(k_{j})+\bar{P}(k_{i},k_{j}) \biggl) \notag \\
& + \frac{1}{\bar{n}V}\biggl(4\bar{B}(k_{i},k_{j})+\bar{B}(0,k_{j})+\bar{B}(k_{i}, 0) \biggl)  \notag \\
& + \frac{\bar{T}(k_{i},k_{j})}{V} + \frac{(1+\alpha^{3})}{\bar{n}^{3}V}.
\label{eq:covnowin}
\end{align}
We have denoted this covariance matrix $\bC^{sm}$ to distinguish it from the full covariance matrix in the presence of a window function, the expression for which is given in Appendix~\ref{sec:appcov} and will be visited later. $V$ is the volume of the simulation, whilst $\bar{n}$ is the number density of tracers, which must be constant by definition in the absence of a window function. $\alpha$ is the ratio of tracers to synthetic data points that is used to estimate the clustering of the field. $\bar{P}$, $\bar{B}$ and $\bar{T}$ are the bin-averaged power spectrum, bispectrum and trispectrum,
\begin{align}
\bar{T}(k_{i},k_{j}) &= \int_{V_{k_{i}}} \frac{d^{3}k}{V_{k_{i}}}\int_{V_{k_{j}}} \frac{d^{3}k '}{V_{k_{j}}}T(\bk,\bk ',-\bk, -\bk'), \\
\bar{B}(k_{i},k_{j}) &= \int_{V_{k_{i}}} \frac{d^{3}k}{V_{k_{i}}}\int_{V_{k_{j}}} \frac{d^{3}k '}{V_{k_{j}}}B(\bk,\bk ',-\bk-\bk'), \\
\bar{P}(k_{i},k_{j}) &= \int_{V_{k_{i}}} \frac{d^{3}k}{V_{k_{i}}}\int_{V_{k_{j}}} \frac{d^{3}k '}{V_{k_{j}}}P(|\bk+\bk '|), \\
\bar{P}(k_{i}) &= \int_{V_{k_{i}}} \frac{d^{3}k}{V_{k_{i}}}P(\bk),
\end{align}
where the two-, three- and four-point functions for each mode $\bk$ and $\bk'$ are averaged over k-space volumes $V_{k_{i}}$ and $V_{k_{j}}$.

Hence the error on the covariance matrix measured from a set of mock catalogues is inversely proportional to the volume of a survey being simulated. Knowledge of this behaviour can be used to augment the standard method of estimating the covariance matrix from simulations, and improve the error on the covariance matrix given a fixed computational time. The known scaling of the covariance matrix means we can run simulations of smaller size than required to fit a survey, measure their covariance and then scale it by the appropriate volume to the covariance that would have been measured from a set of simulations large enough to contain the survey volume. Running smaller volume simulations means that more simulations can be run in a fixed time, and hence the error on the estimate of the covariance improves. This also has the additional benefit that each simulation will be easier to run in terms of memory consumption and could be made more accurate in terms of the non-linear physics. 

\subsection{A demonstration with Gaussian Random Fields}

As a simple proof of concept, take Eq.~\ref{eq:coverr} and the case of a set of $N_{L}$ large simulations, with volume $V_{L}$. The error on the covariance matrix measured from those simulations is 
\begin{equation}
\Delta\hat{\bC}_{L} \propto \sqrt{\frac{2}{N_{L}}}\frac{1}{V_{L}}.
\label{eq:errbegin}
\end{equation}

Now take a set of twice as many smaller simulations $N_{S} = 2N_{L}$, each half the volume of the larger simulations, $V_{S} = 1/2V_{L}$. Naively one would expect running this set to take the same amount of computational time as the larger volume set (in reality it would be even less due to the imperfect scaling of most simulation codes). The error on the covariance matrix measured from these would be
\begin{equation}
\Delta\hat{\bC}_{S} \propto \sqrt{\frac{2}{N_{S}}}\frac{1}{V_{S}} \propto \sqrt{2}\Delta\hat{\bC}_{L}.
\end{equation}
The error using the small volume mocks is actually larger than using the larger volume mocks. This is because the four-point nature of the covariance matrix means that volume is more important than number of simulations. Doubling the volume adds twice as many modes available for estimating the covariance compared to doubling the number of simulations. This is unlike the error on the power spectrum averaged over many simulations, which is two-point in nature and so the doubling the volume has the same effect on this as doubling the number of simulations.

However, what we can do is scale the covariance matrix of the small simulations by the volume ratio, adding in information from our knowledge of the analytic behaviour of the covariance matrix, thus
\begin{equation}
\Delta\hat{\bC}_{S,scaled} \propto \sqrt{\frac{2}{N_{S}}}\frac{1}{V_{S}}\frac{V_{S}}{V_{L}} \propto \frac{\Delta\hat{\bC}_{L}}{\sqrt{2}}.
\label{eq:errend}
\end{equation}
Hence the error on the covariance matrix is decreased by the square root of the number of additional simulations that can be run in the same time period.

To test this scaling we use a set of Gaussian Random Fields (GRFs) based an initial power spectrum generated using {\sc camb} \citep{Lewis2000, Howlett2012}. Each GRF is generated on a Fourier grid with the real and imaginary parts of each Fourier mode $\delta_{\bk}$, drawn from a distribution with variance given by the input dimensionless power spectrum $\Delta^{2}_{\bk} = |\bk|^{3}P(\bk)/2\pi^{2}$, i.e., 
\begin{equation}
\mathcal{P}(\delta_{\bk}) = \frac{1}{\sqrt{2\pi \Delta^{2}_{\bk}}}e^{-\frac{\delta^{2}_{\bk}}{2\Delta^{2}_{\bk}}}.
\end{equation}

500 GRFs were generated on a grid of edge-length $L=1280\mpcoh$ consisting of $512^{3}$ cells, whilst 4000 were generated on a grid of edge-length $L=640\mpcoh$ consisting of $256^{3}$ cells. Hence the volume of the larger GRFs is 8 times that of the smaller set, however there are 8 times fewer. The power spectrum and covariance matrix from each set was then calculated in bins of width $\Delta k=0.01\hompc$ and $\Delta k=0.04\hompc$.

\begin{figure}
\centering
\includegraphics[width=0.5\textwidth]{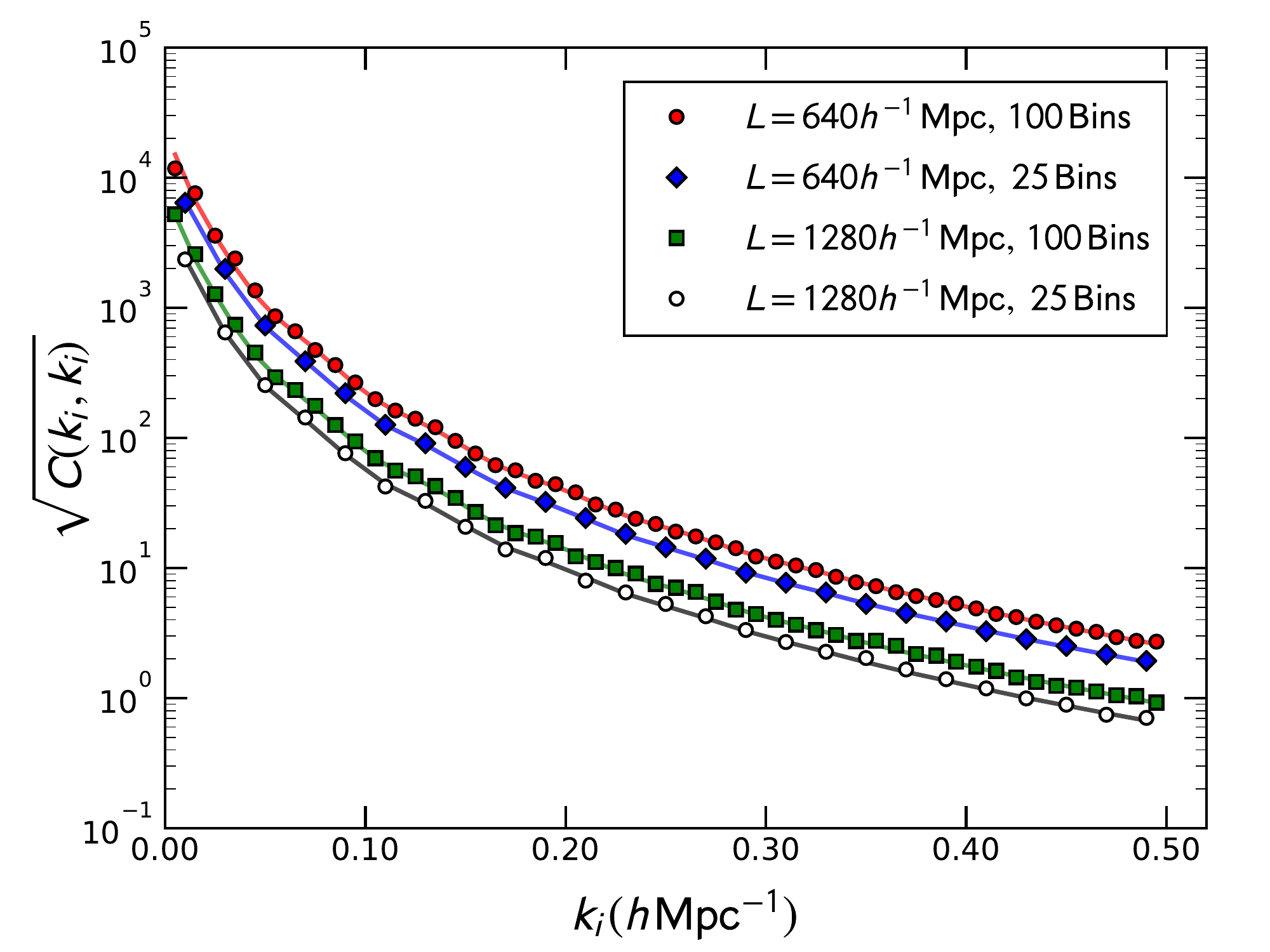}
  \caption{The error on the power spectrum from the two sets of Gaussian Random Fields described in the text with different volumes and measurement bin widths. Points denote the measurements whilst the solid lines show the theoretical predictions. Increasing the bin width and the volume decreases the covariance as there are more modes in each bin to average over.}
  \label{fig:gaussianvolume}
\end{figure}

In the Gaussian regime with no shot-noise, only the term proportional to the power spectrum squared remains in Eq.~\ref{eq:covnowin}. For the two sets of GRFs, the measured variance should match this analytic prediction exactly. This is shown in Fig.~\ref{fig:gaussianvolume}. The agreement between the two is exact within the limits of noise in the measured covariance matrix arising from using a finite number of realisations.

\begin{figure}
\centering
\includegraphics[width=0.5\textwidth]{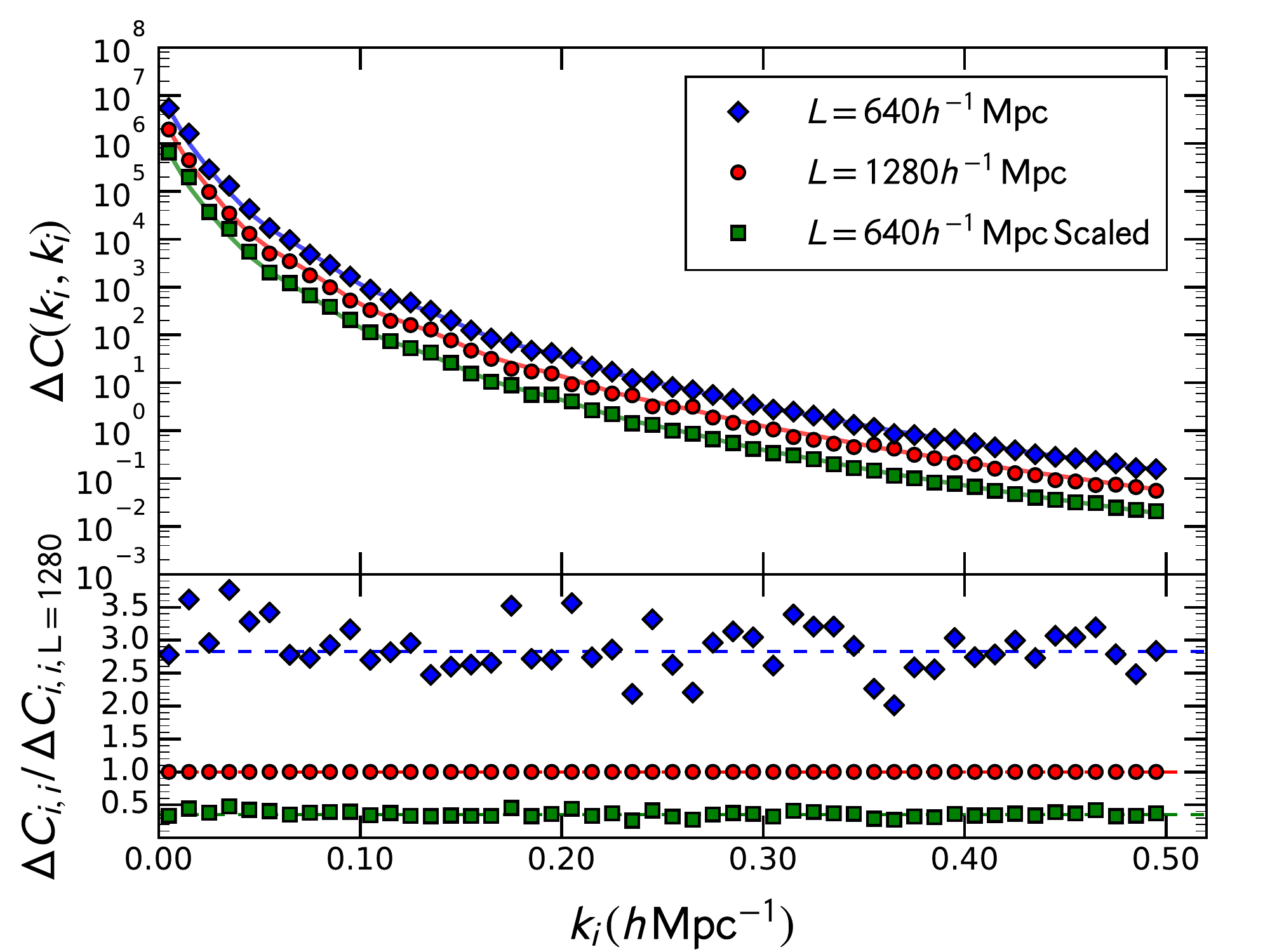}
  \caption{The error on the variance from the two sets of Gaussian Random Fields described in the text with different volumes and including the volume scaling of the small volume simulations. Points show the measurements, whilst solid lines show the theoretical expectation for a GRF (including scaling), Eqs~\ref{eq:coverr} and \ref{eq:errbegin}-\ref{eq:errend}. Shown in the bottom panel is the ratio of the errors compared to the errors in the large volume simulations, compared with the expectations from Eq.~\ref{eq:errbegin}-~\ref{eq:errend}. Generally, decreasing the volume and increasing the number of simulations in concordance increases the covariance matrix error as volume is more important than number of simulations due to their effects on the number of modes available for measuring the four-point nature of the covariance matrix. However, using the volume dependence of the covariance matrix allows this to be counteracted, and can cause the error on the smaller simulations to improve compared to the larger simulations as more simulations can be run in a fixed computational time.}
  \label{fig:gaussianerr}
\end{figure}

The error on the two covariance matrices from the $\Delta k=0.01\hompc$ simulation set was then calculated using bootstrap resampling with replacement over the 500 (4000) large (small) volume GRFs. The error was also calculated where the covariance matrix for each bootstrap sample was scaled by the volume ratio between the large and small simulations, using the analytic behaviour of the covariance matrix to reduce the error. The standard deviation of the three different covariance matrices are shown in Fig.~\ref{fig:gaussianerr}. Also shown in the ratio between the standard deviations of the covariances matrices with that measured from the larger volume GRFs. We expect that the error on the smaller volume simulations will be a factor of $\sqrt{8}$ larger than that of the larger volume simulations, even though there is a factor of 8 more of them as the reduction in volume outweighs the extra simulations. This is indeed seen in Fig.~\ref{fig:gaussianerr}. However when we include the volume scaling of the covariance matrix, the error improves by a factor $\sqrt{8}$ in the small simulations compared to larger simulations, validating Eq.~\ref{eq:errend}.

\subsection{Money for nothing?}

Perhaps this seems to good to be true. We seem to be gaining information at no cost and if so, why can we not make each simulation infinitesimally small? The answer is that the information we are gaining comes from our knowledge of the analytic behaviour the covariance matrix. However, we cannot make each simulation infinitesimally small as the volume scaling of the covariance matrix breaks down as the size of the simulation approaches the scales of interest in the power spectrum.

In particular, the lack of large scale modes in smaller volume simulations has an effect on the covariance matrix measured on both large and scales. On large scales, there must still be enough modes that the power spectrum and its covariance matrix can be measured. Additionally, it has been well documented in recent literature that the presence of coupling between long (on the order of the simulation size) and small scale modes increases the covariance on small scales \citep{Takada2013, Li.Y2014a}. Hence a simulation of a given volume will not return the `true' small scale covariance due to the absence of modes larger than the simulation box.

Finally, for the rescaling method to be viable we must also find a new way to account for the window function of a survey, which can no longer be included by simply cutting out the survey mask from the simulation. In the remainder of this paper we will cover new methods to include larger-than-box modes and the effects of a window function before bringing everything together and showing that we can recover the covariance matrix measured from a set of realistic, traditional galaxy mocks but using simulations only 1/8th the size.

\section{Supersample Covariance} \label{sec:supersamp}
The non-linear nature of gravitational evolution intimately couples long and short wavelength density fluctuations. Due to the cosmological principle, on the very largest scales the density fluctuations should tend to zero. However, when observations of the universe are made, the finite size of the survey means that there may be density fluctutations larger than the survey that couple with modes inside the survey. Though these long wavelength perturbations cannot be measured directly, their interaction with the sub-survey modes still leaves additional information within the covariance matrix. This additional information is commonly known as beat-coupling, halo sample variance or, as will be adopted here, supersample covariance.

The effect of super-survey modes on the power spectrum covariance matrix was originally studied by \cite{Hamilton2006} and \cite{Rimes2006}. \cite{Hu2003} also investigated the effect of these modes on the number counts of halos. Since then there have been many investigations into the nature of supersample covariance as well as its, possibly measurable, information content \citep{Sefusatti2006,Takada2007,Sato2009,Takada2009,Takahashi2009,DePutter2012,Kayo2013,Takada2013,Li.Y2014a,Li.Y2014b,Takahashi2014}.

\cite{Takada2013} give a detailed mathematical description of supersample covariance and its origin. In their work they find that supersample covariance arises from the response of the power spectrum to a rescaling of the background by a long-wavelength mode, which in turn can be related to a particular trispectrum configuration. In this configuration, the quadrilaterals that make up the trispectrum consist of two, nearly equal and opposite, long wavelength modes, $\bq_{12}$. The two orthogonal modes $\bk$ and $\bk'$, are small, and so the trispectrum acts as the modulation of two short wavelength power spectra $P(k)$ and $P(k')$ by some background mode $\delta_{b}$. This is related to the peak-background split framework \citep{Kaiser1984,Cole1989}, in which large-scale galaxy bias can be understood by considering that a long wavelength density perturbation modulates the amplitude of small scale pairs and changes the relative abundances of local peaks above the collapse threshold. Mathematically, the clustering quantity of interest is 
\begin{multline}
T(\bk,-\bk+\bq_{12},\bk',-\bk'-\bq_{12}) \\
\approx T(\bk,-\bk,\bk',-\bk') + \frac{\partial P(k)}{\partial \delta_{b}} \frac{\partial P(k')}{\partial \delta_{b}} P^{L}(q_{12}).
\end{multline}
As the mode $\bq_{12}$ has a long wavelength, the power spectrum of this mode is the linear power spectrum, $P^{L}(q_{12})$.

Using the above expression for the trispectrum in the covariance matrix results in a modified expression for the small-scale covariance that would be measured within the same volume, but which includes the effects of modes larger than the survey
\begin{equation}
\mathsf{C}^{ssc}(k_{i},k_{j}) = \mathsf{C}^{sm}(k_{i},k_{j})+\sigma_{b}^{2}\frac{\partial P(k_{i})}{\partial \delta_{b}} \frac{\partial P(k_{j})}{\partial \delta_{b}},
\label{eq:covssc}
\end{equation}
where
\begin{equation}
\sigma_{b}^{2} = \int \frac{d^{3}k}{(2\pi)^{3}} |W(\bk)|^{2} P^{L}(k)
\label{eq:sigmab}
\end{equation}
is the variance of the background mode $\delta_{b}$ within some window $W(\bk)$. 

The above equation assumes that the density fluctuations are defined with respect to the \textit{global} mean density, $\bar{\rho}_{m}$. In the context of large scale structure analyses, we instead usually estimate the overdensity with respect to the mean density within the local survey volume, $\bar{\rho}^{\mathrm{loc}}_{m}$. Compared to the global mean density, the mean density within the survey volume is modulated by the same background mode that gives rise to the supersample covariance, such that
\begin{equation}
\bar{\rho}^{\mathrm{loc}}_{m} = (1+\delta_{b})\bar{\rho}_{m}
\label{eq:svbox}
\end{equation}
and our estimate of the Fourier space overdensity referenced to the local mean $\delta^{\mathrm{loc}}(\bk)$, is related to the true overdensity via \citep{DePutter2012}  
\begin{equation}
\delta^{\mathrm{loc}}(\bk) = \delta(\bk)/(1+\delta_{b}).
\end{equation}
Strictly speaking, the mean density within the survey enters into both the numerator and denominator when we compute the overdensity which introduces an additional term $-\delta_{b}/(1+\delta_{b})$ into the real-space overdensity measured within the survey volume. However, this constant term disappears for $\bk\neq0$ when we take the Fourier transform of the overdensity field. From the overdensity referenced to local means the measured power spectrum becomes 
\begin{equation}
P^{\mathrm{loc}}(k) = P(k)/(1+\delta_{b})^{2}
\label{eq:pkglob}
\end{equation}
and it is the variation of $P^{\mathrm{loc}}(k)$ with $\delta_b$ that we are interested in for the supersample covariance term in the measured covariance matrix. Thus the revised covariance referenced to local means is
\begin{multline}
\mathsf{C}^{ssc,\mathrm{loc}}(k_{i},k_{j}) = \mathsf{C}^{sm}(k_{i},k_{j})\\
+\sigma_{b}^{2}\left(\frac{\partial P(k_{i})}{\partial \delta_{b}}-2P(k_{i})\right)\left(\frac{\partial P(k_{j})}{\partial \delta_{b}}-2P(k_{j})\right).
\label{eq:covsscloc}
\end{multline}
and $\sigma_{b}$ is the same as that given in Eq.~\ref{eq:sigmab}.

The formalism of \cite{Takada2013} provides a useful way of characterising the effect of supersample covariance on cosmological measurements and of disentangling and utilising the signal from modes outside the survey in obtaining cosmological constraints \citep{Li.Y2014b}. Of direct interest to this study however is the work of \cite{Li.Y2014a} who detail the effect of supersample covariance on \textit{simulations}.

Unlike in surveys, where modes outside the volume encode information inside the volume, periodic simulations have no external modes. These are implicitly set to zero along with the average overdensity. Hence the covariance measured from an ensemble of simulations will be lower than that measured from an ensemble of real surveys of the same volume. Similarly the covariance of a set of small volume simulations will be lower than that of a sub-volumes drawn from a larger set of simulations (even after scaling by the volume) due to the absence of modes larger than the small volume. Some of these are present in the large volume simulation. However, on top of this the large volume simulation will itself be missing modes that would be present in an even larger simulation, though the effect of super-survey modes will diminish as larger and larger volumes are simulated. 

Hence an estimate of the true covariance for some measured survey requires the inclusion of modes larger than the simulation volume. This is identified in \cite{Li.Y2014a} who find that a set of small volume simulations can significantly underestimate the covariance even on moderately large ($k\approx0.1$) scales. They also investigate analytic methods of including modes larger than the simulation volume. If the scaling method presented within this paper is to work effectively, a method for introducing `larger than box' modes into the small volume simulations will also have to be included.

\subsection{Computing supersample covariance using the Separate Universe approach}\label{sec:separateuniverses}

In this work we present two methods for including supersample covariance in simulations that still allows us to `volume-scale' the small volume covariance matrix. Both of these methods are based on the separate universe approach of \cite{Sirko2005} (also presented by \citealt{Baldauf2016}), but differ in how the additional covariance is computed and applied to the small volume covariance matrix.

In the separate universe approach, the background mode is treated as a density contrast which is then absorbed into the mean density of the simulation as in Eq.~\ref{eq:svbox}, where $\bar{\rho}^{\mathrm{loc}}_{m}$ is now the effective mean density of the simulation and $\bar{\rho}_{m}$ is the mean density given the fiducial cosmological parameters.

\cite{Sirko2005} shows that this change in the mean density for each simulation can be modelled by modifying the input cosmology used to run each simulation via the parameterisation

\begin{equation}
\begin{aligned}
a_{\mathrm{box}} &= a\biggl(1-\frac{D(a)\delta_{b,0}}{3D(1)}\biggl), \\
H_{0,\mathrm{box}} &= H_{0}(1+\phi)^{-1}, \\
\Omega_{m,0,\mathrm{box}} &= \Omega_{m,0}(1+\phi)^{2}, \\
\Omega_{\Lambda,0,\mathrm{box}} &= \Omega_{\Lambda,0}(1+\phi)^{2}, \\
\Omega_{k,0,\mathrm{box}} &= 1-(1+\phi)^{2}(\Omega_{m,0}+\Omega_{\Lambda,0}),
\end{aligned}
\end{equation}
where
\begin{equation}
\phi = \frac{5\Omega_{m,0}}{6}\frac{\delta_{b,0}}{D(1)},
\end{equation}
$\delta_{b,0}$ is the background mode at redshift 0, $D$ is the linear growth factor, $a$, $H_{0}$, $\Omega_{m,0}$, $\Omega_{\Lambda,0}$, $\Omega_{k,0}$ define the output scale factor and cosmology of the ensemble, and $a_{\mathrm{box}}$, $H_{0,\mathrm{box}}$, $\Omega_{m,0,\mathrm{box}}$, $\Omega_{\Lambda,0,\mathrm{box}}$, $\Omega_{k,0,\mathrm{box}}$ are the parameters given to each realisation.

Because the scale factors, $a_{\mathrm{box}}$ are different for each simulation, the physical scale of each simulation is different. In order to simplify the covariance calculation and match modes within bins, it is advantageous for the physical scale of each simulation to coincide at their respective output times. To do this we can modify the size of each simulation to be
\begin{equation}
L_{\mathrm{box}} = L \frac{a}{a_{\mathrm{box}}}\frac{H_{0,\mathrm{box}}}{H_{0}}
\end{equation}
where $L$ is the size of the unmodified simulation. To emphasise, modifying the box size in this way does not account for the effects of larger-than-box modes; this requires us to modify the effective density in the simulation, which is done by changing $a$, $\Omega_{m}$ and the other cosmological parameters. Rather, changing the box size just allows us to easily compare modes between boxes output at different scale factors.

Given the separate universe prescription, we present our two methods for including supersample covariance in volume-scaled simulations below. Our goal is to recover the covariance matrix $\bC^{surv}$, that would be measured in a \textit{survey} with volume $V^{surv}$ and includes the effects of modes outside the survey volume. The formalism of \cite{Takada2013} demonstrates how this could be done, but is only valid if the volume of the cubic simulations, $V^{sm}$ used to calculate $\bC^{sm}$ is equal to the survey volume, in which case $\bC^{surv}$ = $\bC^{ssc}$ in Eqs.~\ref{eq:covssc} and similarly for Eq~\ref{eq:covsscloc}. We instead aim to do this with cubic simulations with a volume that may not be equal to $V^{surv}$ (and, to reduce computational requirements, is ideally much smaller). The most obvious way to do this is by first scaling the simulation covariance matrix to the effective survey volume, then adding on the supersample covariance separately. We call this the `addition' method. The second method, which we find preferable in terms of both accuracy and computational cost, adds the supersample covariance (for the survey volume) directly into the simulations \textit{before} scaling. We call this the `ensemble' method. An important point to remember is that, regardless of the simulation size V$^{sm}$, we need to recover the covariance corresponding to the survey. Hence, if we scale a covariance matrix with no supersample covariance correction by the ratio of the survey and simulation volumes as per Eq.~\ref{eq:covnowin}, we will need to include the supersample covariance corresponding to the survey. This follows because we are directly constructing the covariance matrix for the survey volume. If instead, we were using realisations of the survey drawn from larger simulations, the supersample covariance would depend on the simulation rather than survey volume. In effect, in this case the set of survey realisations would already have a small scatter in background density, and we would have to take care defining and using local, global and simulation mean densities. 

\subsubsection{Addition Method}
Our first method, the `addition' method, relies on using a small number of separate universe simulations to evaluate the supersample covariance term (the second term in Eq.~\ref{eq:covssc}) which is then added to the scaled, small volume covariance matrix. In this case, combining Eqs.~\ref{eq:covnowin} and \ref{eq:covssc}, we can write
\begin{multline}
\mathsf{C}^{surv}(k_{i},k_{j}) = \frac{V^{sm}}{V^{surv}}\mathsf{C}^{sm}(k_{i},k_{j})+\sigma_{b}^{2}\frac{\partial P(k_{i})}{\partial \delta_{b}} \frac{\partial P(k_{j})}{\partial \delta_{b}},
\end{multline}
where $\sigma^{2}_{b}$ is now calculated from Eq.~\ref{eq:sigmab} with the linear power spectrum for the fiducial cosmology and the window function corresponding to $V^{surv}$. We can write a similar expression for $\bC^{surv,\mathrm{loc}}$. In this work, we compute the small volume covariance matrix $\bC^{sm}$ using cubic simulations with \textit{fixed} input parameters and overdensities referenced to the local mean within each small simulation. We then scale this covariance matrix to obtain the covariance matrix without the supersample covariance correction, corresponding to the survey volume. For the supersample covariance term, we calculate the response of the power spectrum to the background modes outside the survey using separate universe simulations and the `growth-dilation' method of \citealt{Li.Y2014a} (their Eq. 47), where we generate pairs of realisations with the cosmology of each pair modified by $\delta_{b}=\pm0.01$. The measured power spectra from each pair is then finite-differenced to obtain the power spectrum response. In principal only a single pair of simulations generated from the same initial conditions but with different $\delta_{b}$ is necessary to compute this, however the realization of small scale power in the separate universe simulations introduces stochasticity in the response calibration, which can be reduced by averaging over multiple realizations. The size of the separate universe simulations used to calculate the supersample covariance term is largely unimportant, as we only need them to calibrate the response of the power spectrum to a background mode, and we know the scaling of the supersample covariance correction. For convenience, we use separate universe simulations with volume $V^{sm}$. For any application of the `addition’ method we will always require more simulations than our second method due to the fact that we need both an estimate of the small volume covariance and multiple separate universe realisations.

\subsubsection{Ensemble Method}
To remove the need to evaluate the supersample covariance term separately (and hence require no extra simulations), we develop a second method which incorporates the separate universe approach directly into the the ensemble of small volume simulations in a way that recovers the supersample covariance corresponding to the survey volume. We begin with the ansatz that as the background mode present in any survey is a large scale mode, it is expected to be drawn from a Gaussian distribution with variance $(\sigma_{b})^{2}$.

Based on this, we can include supersample covariance in a set of simulations by doing the following:
\begin{enumerate}
\item{Calculate $\sigma_{b}$ based on the input linear power spectrum at redshift zero and the survey window function.}
\item{For each simulation draw a background mode $\delta_{b,0}$ from a Gaussian distribution with zero mean and variance given by $V^{surv}\sigma_{b}^{2}/V^{sm}$.}
\item{Evaluate the new cosmology, output redshift and boxsize for each simulation, based on the values of $\delta_{b,0}$.}
\item{Run the simulations as normal, but compute the particle positions and power spectra in box coordinates, i.e., with the box length $L_{\mathrm{box}}$ scaled out of the particle positions.}
\item{Finally, with the power spectra in box coordinates, calculate the covariance matrix as normal. When comparing length scales between the `ensemble' method and the survey (sub-volume) covariance matrix in the following section, we simply multiply by $L$ to convert from box coordinates. We denote the covariance matrix measured using this method $\bC^{sm,\mathrm{loc}}_{\delta_{b}}$.}
\end{enumerate}

Following this procedure means that the covariance matrix evaluated from the modified small volume simulations can be written
\begin{align}
&\mathsf{C}^{sm,\mathrm{loc}}_{\delta_{b}}(k_{i},k_{j}) = \mathsf{C}^{sm}(k_{i},k_{j})+\frac{V^{surv}}{V^{sm}}\sigma_{b}^{2} \nonumber \\
&\qquad\left(\frac{\partial P(k_{i})}{\partial \delta_{b}}-2P(k_{i})\right)\left(\frac{\partial P(k_{j})}{\partial \delta_{b}}-2P(k_{j})\right) \nonumber\\
&\qquad \qquad \qquad =\frac{V^{surv}}{V^{sm}}\mathsf{C}^{surv,\mathrm{loc}}(k_{i},k_{j})
\end{align}
and our end-goal of $\bC^{surv,\mathrm{loc}}$ can be recovered simply by multiplying the covariance by the ratio of survey and simulation volumes.

An important point is that the variance of the Gaussian distribution we draw our background modes from is given by $V^{surv}\sigma_{b}^{2}/V^{sm}$ such that we recover the correct contribution to the covariance matrix from supersample modes after scaling. We also modify the box size of each simulation and then run our simulations and compute the power spectra in box coordinates, so that when we compute the covariance matrix, we are comparing the same physical scales. Again, it is the change in cosmology that introduces the supersample covariance. Changing the box size and working in box coordinates just allows us to compute the covariance matrix in the same fashion as the unmodified small volume simulations. 

Finally, if we use the standard approach for computing the power spectrum from simulations and evaluate the mean using the (constant) number of particles in the small volume, we end up with a scaled version of the sub-volume covariance matrix referenced to \textit{local} means. In general, for large scale structure analyses, this is the covariance matrix we are interested in. However, to recover the covariance matrix referenced to global means we can simply multiply the power spectrum of each small volume realisation by $(1+\delta_{b})^{2}$, using the value of $\delta_{b}$ corresponding to that realisation, before computing the covariance matrix (see Eq.~\ref{eq:pkglob}).

\subsection{Tests on {\sc l-picola} Simulations} \label{sec:lpicolassc}

The two methods given in the previous section should work for any simulation code, although care must be taken to ensure that all parameters that depend on the background mode are modified correctly. In order to demonstrate their effectiveness we use a set of fast, non-linear dark matter simulations generated using the approximate N-Body code {\sc l-picola} \citep{Howlett2015b}. It was shown in \cite{Howlett2015b} that this code is able to reproduce the clustering of dark matter extremely well on non-linear scales compared to a full N-Body simulation, but at significantly reduced computational cost, which allows for large ensembles to be run easily. In any case as this test is comparative in nature, (we are comparing sets of simulations run using the same code), the choice of simulation code is unimportant.

However, when using {\sc l-picola} for studying the effect of super-survey modes the value of $\sigma_{8}$ that is passed to {\sc l-picola} must also be modified. In the separate universe approach, one would expect that the simulations should be coincident at high redshift. To then ensure that this is true, it is necessary to scale the value of $\sigma_{8}$ that is given to each {\sc l-picola} run. The reason for this is \textit{not} physical; the change in the growth of structure in each simulation has already been captured by the modifications to the input cosmology and output redshift. Rather this is due to the fact that {\sc l-picola} requires an input power spectrum \textit{at redshift zero} and an associated value of $\sigma_{8}$ \textit{at redshift zero} to generate the initial conditions. The code then re-normalises the input power spectrum by the input value of $\sigma_{8}$ internally and the power spectrum at the redshift of the initial conditions is then calculated by scaling the re-normalised redshift zero power spectrum back by the growth factor within the code. 

As the cosmology of each simulation is slightly different, so to is the growth factor. Hence for a fixed input power spectrum and value of $\sigma_{8}$, but different cosmologies, the {\sc l-picola} simulations will not coincide at high redshift. To ensure that they do we can modify $\sigma_{8}$ by the ratio of the normalised growth factors in the fiducial and `box' cosmologies, i.e., 
\begin{equation}
\sigma_{8,\mathrm{box}} = \sigma_{8}\frac{D^{2}(z_{sync},\Omega_{m},\Omega_{\Lambda})}{D^{2}(0,\Omega_{m},\Omega_{\Lambda})}\frac{D^{2}(0,\Omega_{m,\mathrm{box}},\Omega_{\Lambda,\mathrm{box}})}{D^{2}(z_{sync},\Omega_{m,\mathrm{box}},\Omega_{\Lambda,\mathrm{box}})}
\end{equation}
Hence, for every {\sc l-picola} simulation, the input power spectrum is kept fixed, but the code renormalises by $\sigma_{8,\mathrm{box}}$ internally. Given the different cosmologies and growth factors, this then means that the power spectrum as calculated by the code at $z_{sync}$ matches.

The remainder of this section will be dedicated to showing that both the `addition' and `ensemble' methods recover the supersample covariance for a suite of {\sc l-picola} simulations. For this purpose we generate the following suite of simulations. The mass and force resolution for each set is identical.
\begin{itemize}
\item{\textit{Sub-volume/survey}: 500 $L=2048\mpcoh$, $N=1024^{3}$ simulations, where each is split into 8 sub-volumes such that we can use them to calculate the covariance matrix of $L=1024\hompc$ simulations including the effects of supersample covariance, our proxy for $\bC^{surv}$. This is the covariance matrix that our two scaled methods will be compared to and we compute covariance matrices referenced to both global and local means.}
\item{\textit{Small volume}: 4000 $L=256\mpcoh$, $N=128^{3}$ simulations. The cosmological parameters for these simulations are all identical. The covariance matrix from these, $\bC^{sm}$, will be volume scaled by $V^{sm}/V^{surv}=1/64$ to show the covariance matrix for $L=1024\hompc$ simulations \textit{without} supersample covariance.}
\item{\textit{Small volume with $\delta_{b}$}: 4000 $L=256\mpcoh$, $N=128^{3}$ simulations where the cosmology and box size for each simulation has been perturbed by a unique background mode $\delta_{b,0}$. In our second method, the covariance matrix from these, $\bC^{sm,\mathrm{loc}}_{\delta_{b}}$ will also be volume scaled by $V^{sm}/V^{surv}$, but by construction already includes supersample covariance.}
\item{\textit{Separate universes}: $2\times64$ $L=256\mpcoh$, $N=128^{3}$ realisations where each set of 64 has been generated with an identical cosmology corresponding to $\delta_{b,0}=\pm0.01$ and a \textit{fixed} box size of $L_{\mathrm{box}}=256\mpcoh$. These are used to calculate the supersample covariance term separately, or more precisely, the growth term of the power spectrum response. The dilation term is computed using the average power spectrum of the 4000 \textit{small volume} simulations.}
\end{itemize}

\begin{figure*}
\centering
\subfloat{\includegraphics[width=0.5\textwidth]{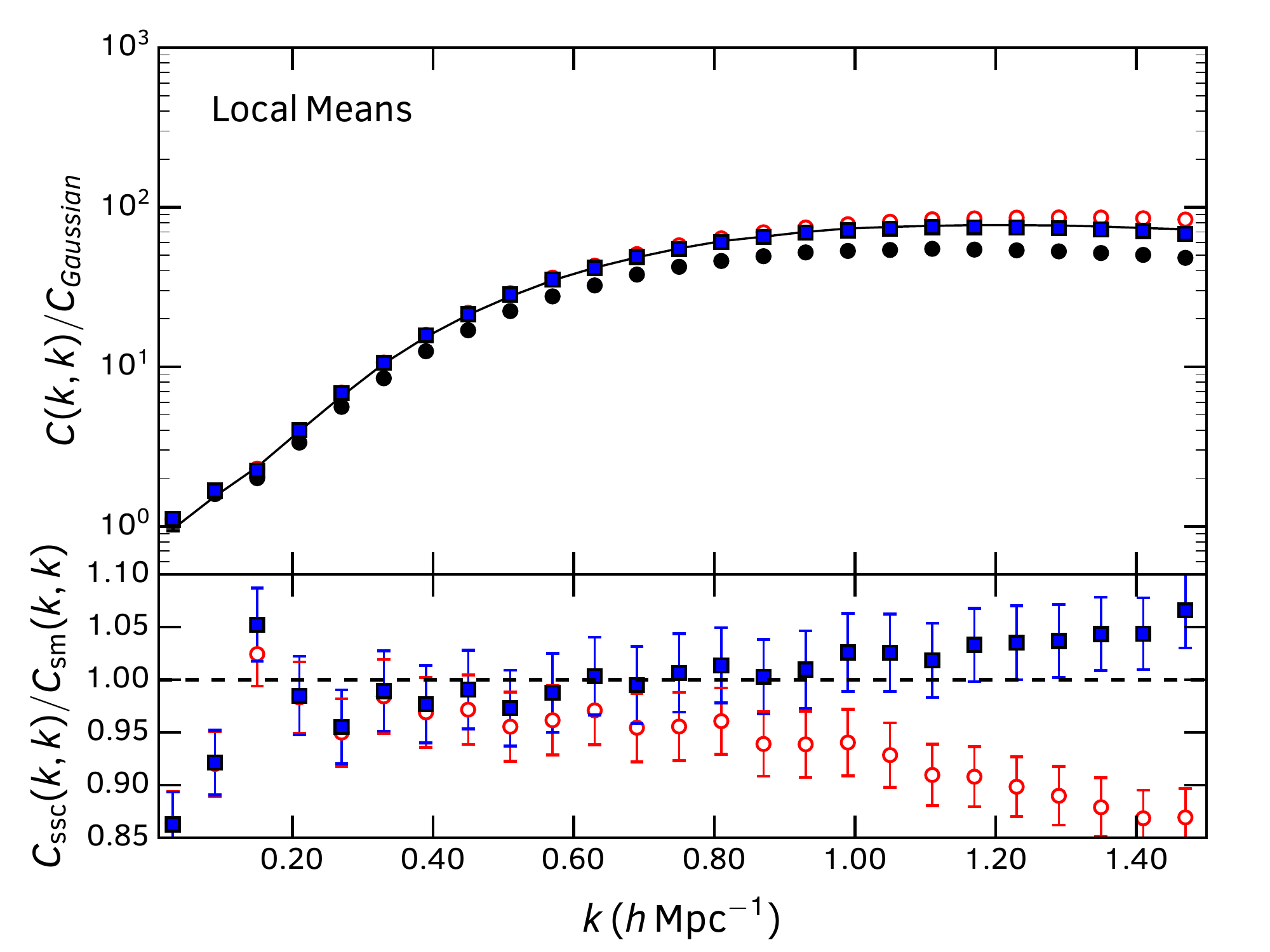}}
\subfloat{\includegraphics[width=0.5\textwidth]{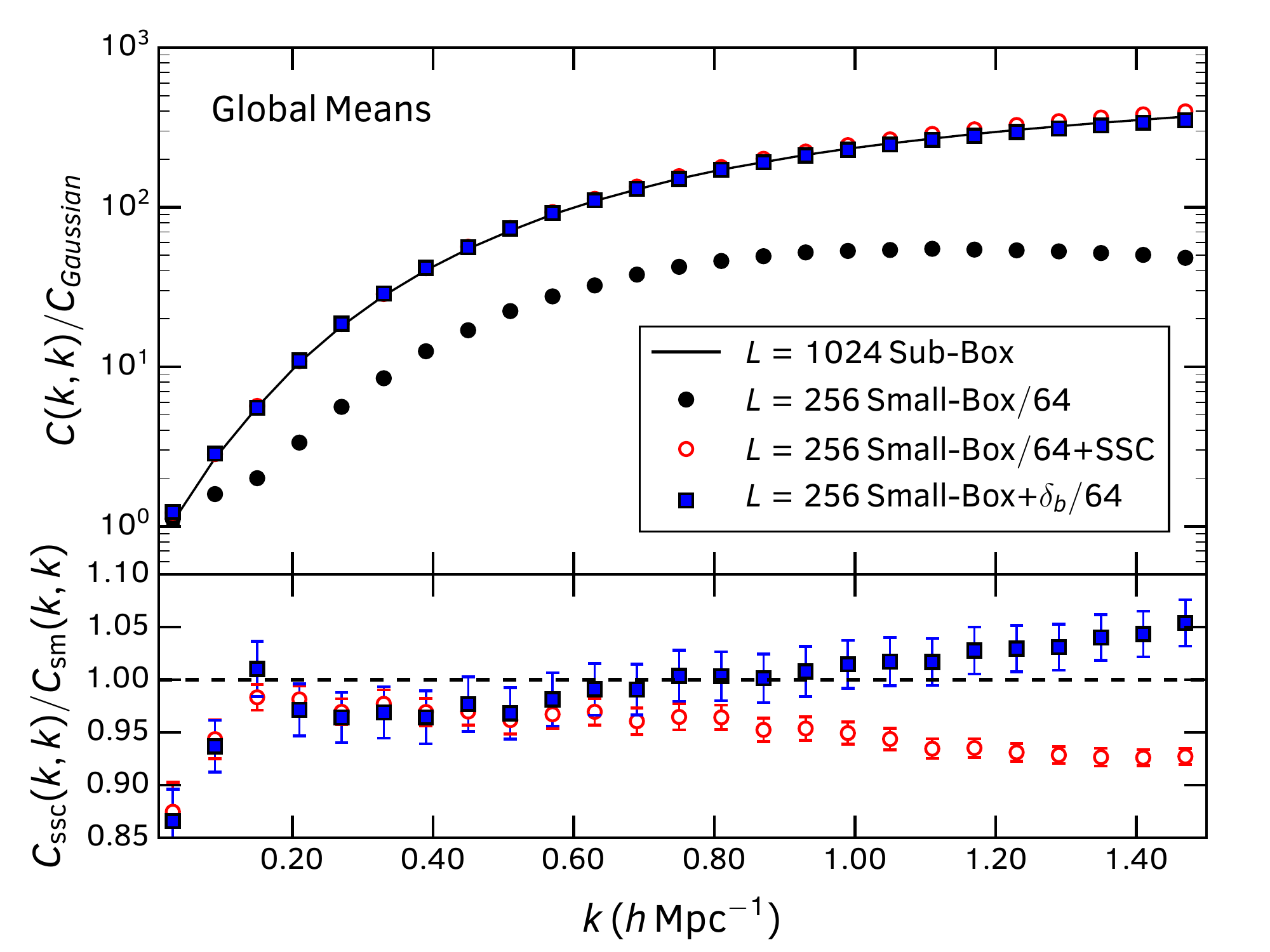}}\\
\subfloat{\includegraphics[width=0.5\textwidth]{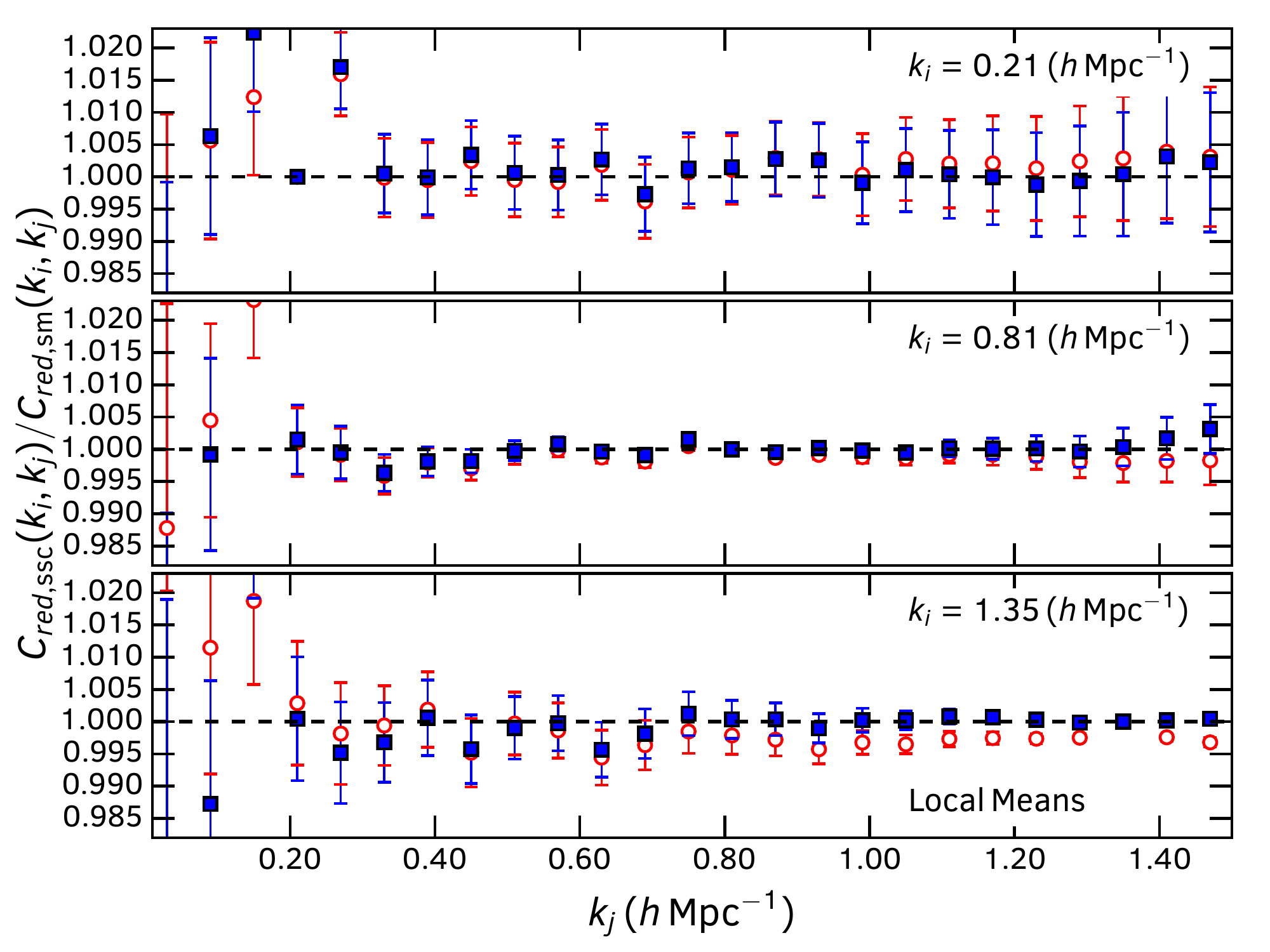}}
\subfloat{\includegraphics[width=0.5\textwidth]{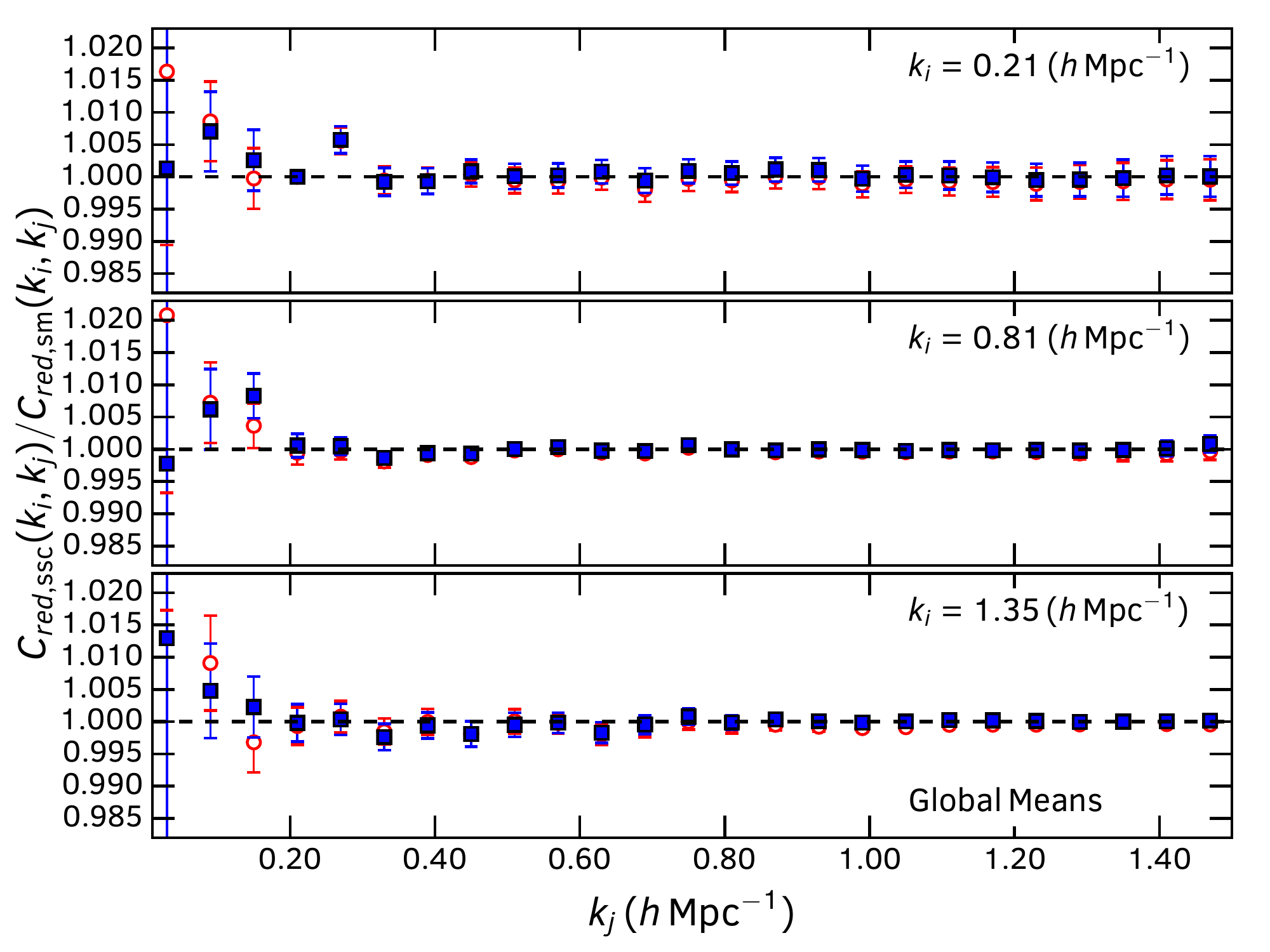}}
\caption{A comparison of the covariance and correlation matrices for $L=1024\mpcoh$ sub-volumes (solid line) against scaled $L=256\mpcoh$ small volume simulations (points) with and without correction for supersample covariance. Black circles, open red circles and blue squares show the scaled small volume covariance matrix without correction, and corrected using the `addition' and `ensemble' methods respectively. Upper panels show the diagonal elements of the covariance matrix normalised by the Gaussian covariance (first term of Eq.~\ref{eq:covnowin}) and the ratio of the scaled small volume covariance against the sub-volume covariance. Lower panels show the ratio between slices of the sub-volume and scaled small volume correlation matrices. We show covariance matrices referenced to local means in the left panels and referenced to global means in the right panels.}
  \label{fig:covssc}
\end{figure*}

Only dark matter simulations are used to test this correction as the non-linear nature of the supersample covariance means that it is largely hidden by shot-noise in a galaxy mock catalogue. All simulations are generated using a linear power spectrum from {\sc camb} and a flat fiducial cosmology with $\Omega_{m}=0.31$, $n_{s}=0.96$ and $\sigma_{8}=0.83$. They are evolved using the modified COLA timestepping method with 11 timesteps from an initial redshift of $z_{i}=9.0$ up to $z_\mathrm{box} = 1.0/a_{\mathrm{box}} - 1.0$. Strictly speaking, we could also modify the initial redshift at which timestepping begins, such that the different realisations spend an equivalent amount of physical time timestepping, however we find negligible difference in the results using $z_{i}$ or $z_{i,\mathrm{box}}$ as the point at which timestepping begins.  

The power spectrum for each simulation is calculated using a number of cells equal to the unmodified length, i.e., a constant cellsize of $1\mpcoh$ in the fiducial cosmology, and the number of cells remains the same between the small volume simulations with and without the background modes, $\delta_{b}$, even though the former is computed in box coordinates. The power spectra and covariance matrices are calculated using 25 bins in the range $0.0<k<1.5\hompc$. The errors on the covariance matrix are calculated using bootstrap resampling.

Figure~\ref{fig:covssc} shows the result of the supersample covariance corrections. We plot the elements and ratios for all four covariance matrices computed from the various simulation sizes and slices through the correlation matrix, $\mathsf{C}_{red}(k_{i},k_{j})=\mathsf{C}(k_{i},k_{j})/\sqrt{\mathsf{C}(k_{i},k_{i})\mathsf{C}(k_{j},k_{j})}$. As expected we find that the presence of supersample modes in the sub-volumes gives a significant increase in the covariance matrix compared to the small volume simulations even after scaling by the volume ratio. This effect is exacerbated when the power spectra are referenced to global means. On large scales, we find that both the `addition' and `ensemble' methods are consistent, and succeed in recovering the supersample covariance. On the largest scales the covariance matrix is overestimated in the small volume simulations due to the lack of modes (the simulations we have used here are significantly smaller than any real large scale structure analysis is likely to use), but generally the ratio between the sub-volume and corrected small volume covariance matrices is accurate to within $5\%$ for $k<1\hompc$. It should be noted that different bins in the covariance matrix will be very highly correlated, so the error bars plotted will not be representative and any residual difference between the sub-volume and corrected small scale covariance matrices may be consistent with noise. We also find that both methods reproduce the correlation matrix in the presence of supersample modes extremely well.

\begin{figure}
\centering
\includegraphics[width=0.5\textwidth]{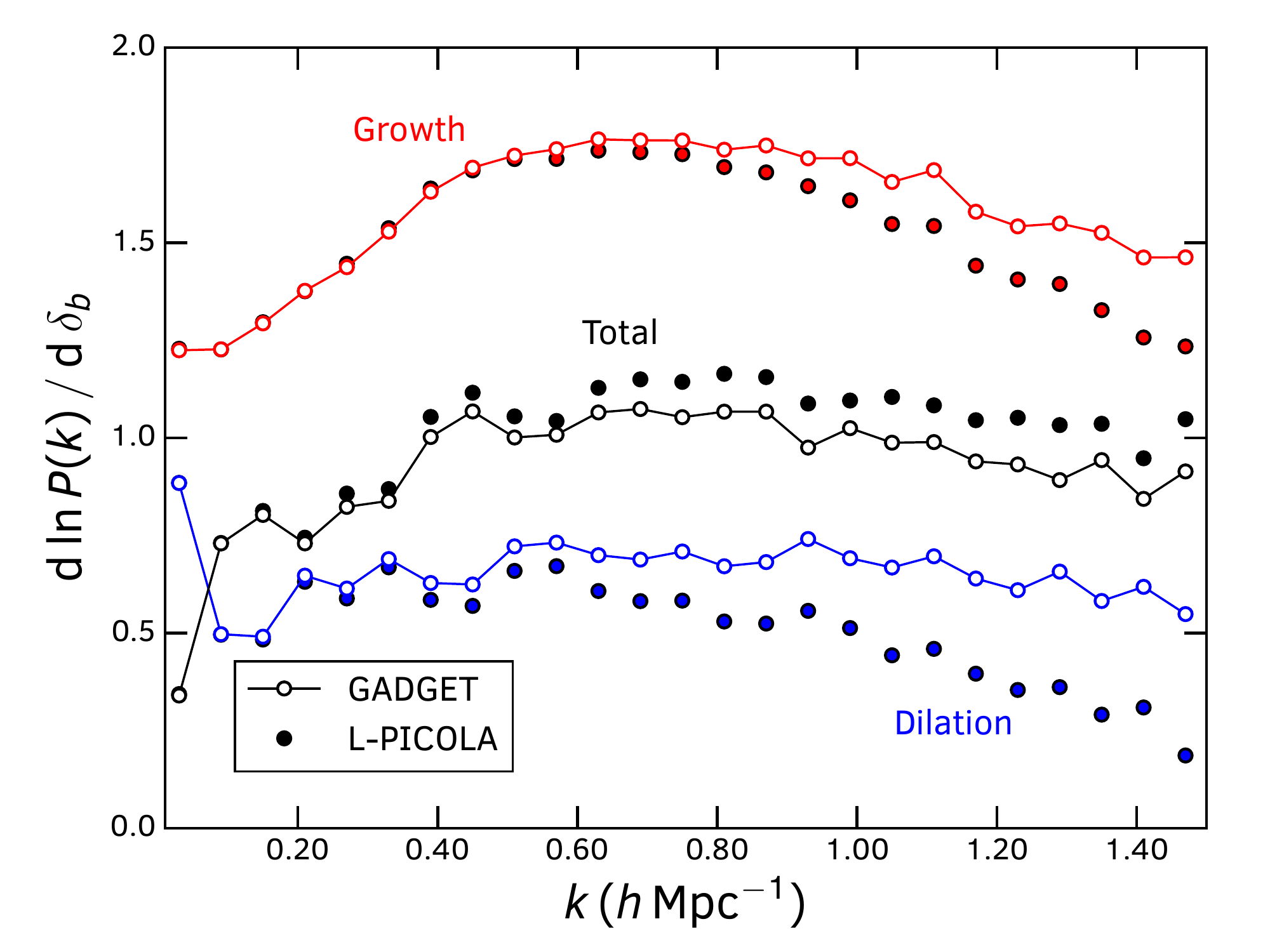}
  \caption{The response of the power spectrum (referenced to local means) to a background mode $\delta_{b}$, separated into growth and dilation terms, calculated using {\sc l-picola} and {\sc gadget-2} simulations started from the same initial conditions. On small scales the {\sc l-picola} simulations underestimate the growth term, but underestimate the dilation term even more which leads to an overestimate of the supersample covariance.}
  \label{fig:growthdilation}
\end{figure}

However, on scales $k>1\hompc$ we find some difference between the diagonal elements of the `addition' and `ensemble' covariance matrices. The ensemble method still agrees within $5\%$, however the `addition' method overestimates the covariance matrix. The cause of this is the use of {\sc l-picola} simulations to calculate the response of the power spectrum to a background mode. Comparing a single set of separate universe simulations (with $\delta_{b}=0$, $\pm0.01$) drawn from identical initial conditions but run with {\sc l-picola} and {\sc gadget-2} \citep{Springel2005}, we find that the use of approximate methods to evaluate this underestimates both the growth and dilation terms, but in such a way that the total response of the power spectrum is overestimated. This in turn causes the supersample covariance to be overestimated. This is shown in Fig.~\ref{fig:growthdilation}. Hence, we conclude that even for power spectra and covariance matrices estimated using approximate dark matter simulations, the response of the power spectrum to a background mode on non-linear scales must be evaluated using accurate N-Body simulations. In terms of computational requirements, this means that the `ensemble' method is preferable as the total number of simulations required is smaller, and approximate simulations can be used for the whole procedure.

\section{Survey window function} \label{sec:window}

In Section~\ref{sec:motivation}, we have shown how reducing the volume of simulations used to measure covariances can improve the errors recovered for a fixed computation time. This relies on knowing the expected scaling of the covariance with volume as in Eq.~(\ref{eq:covnowin}), and will not work if this scaling is complicated by modes outside the simulation volume or a survey window function. We have described a method to correct for the lack of ``supersample'' modes in Section~\ref{sec:supersamp}, here we describe an analytic method to compute the effects of a window function on the covariance matrix.

\subsection{Effects of the window function on the covariance matrix}

\begin{figure*}
\centering
\subfloat{\includegraphics[width=0.5\textwidth]{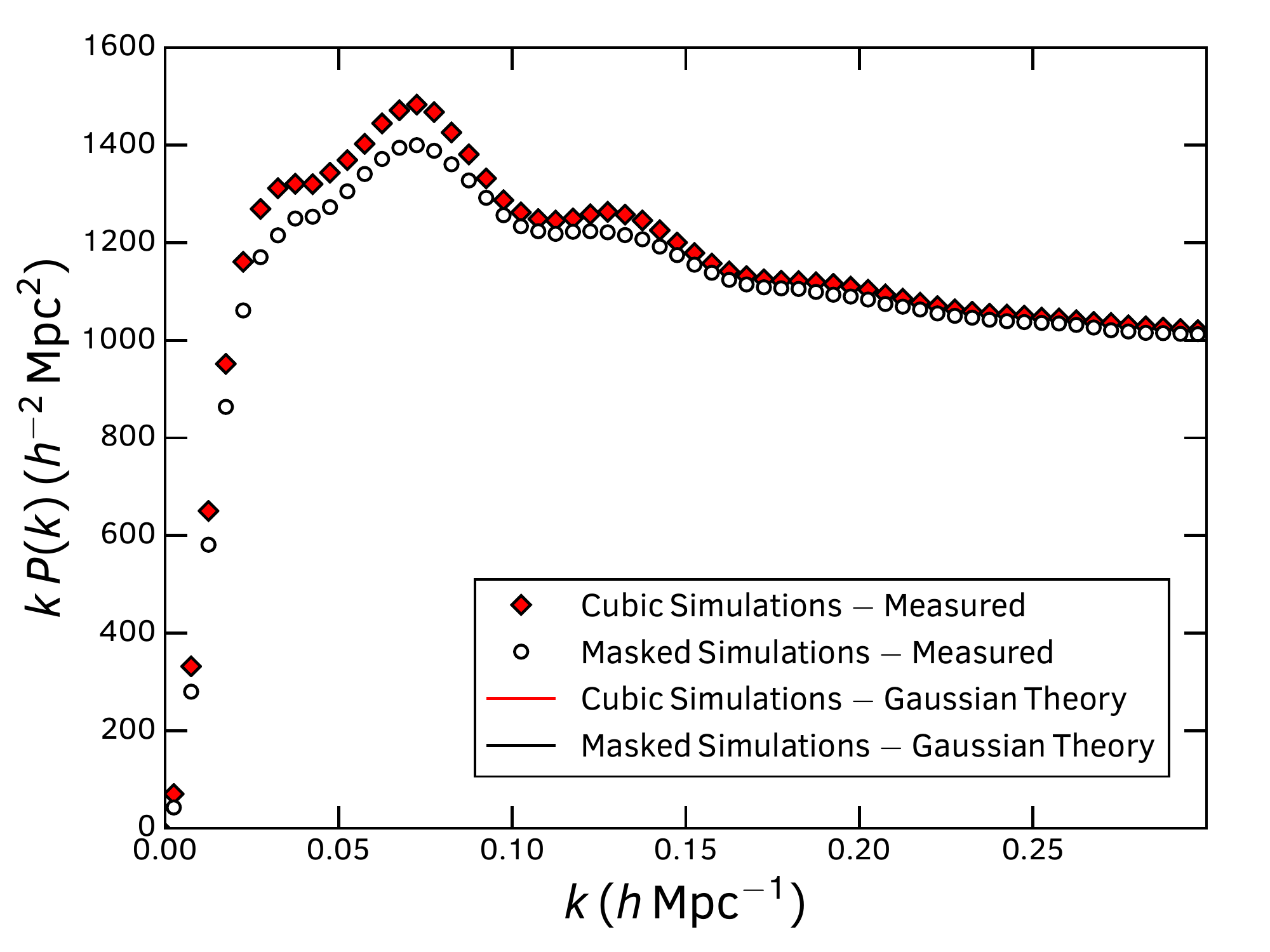}}
\subfloat{\includegraphics[width=0.5\textwidth]{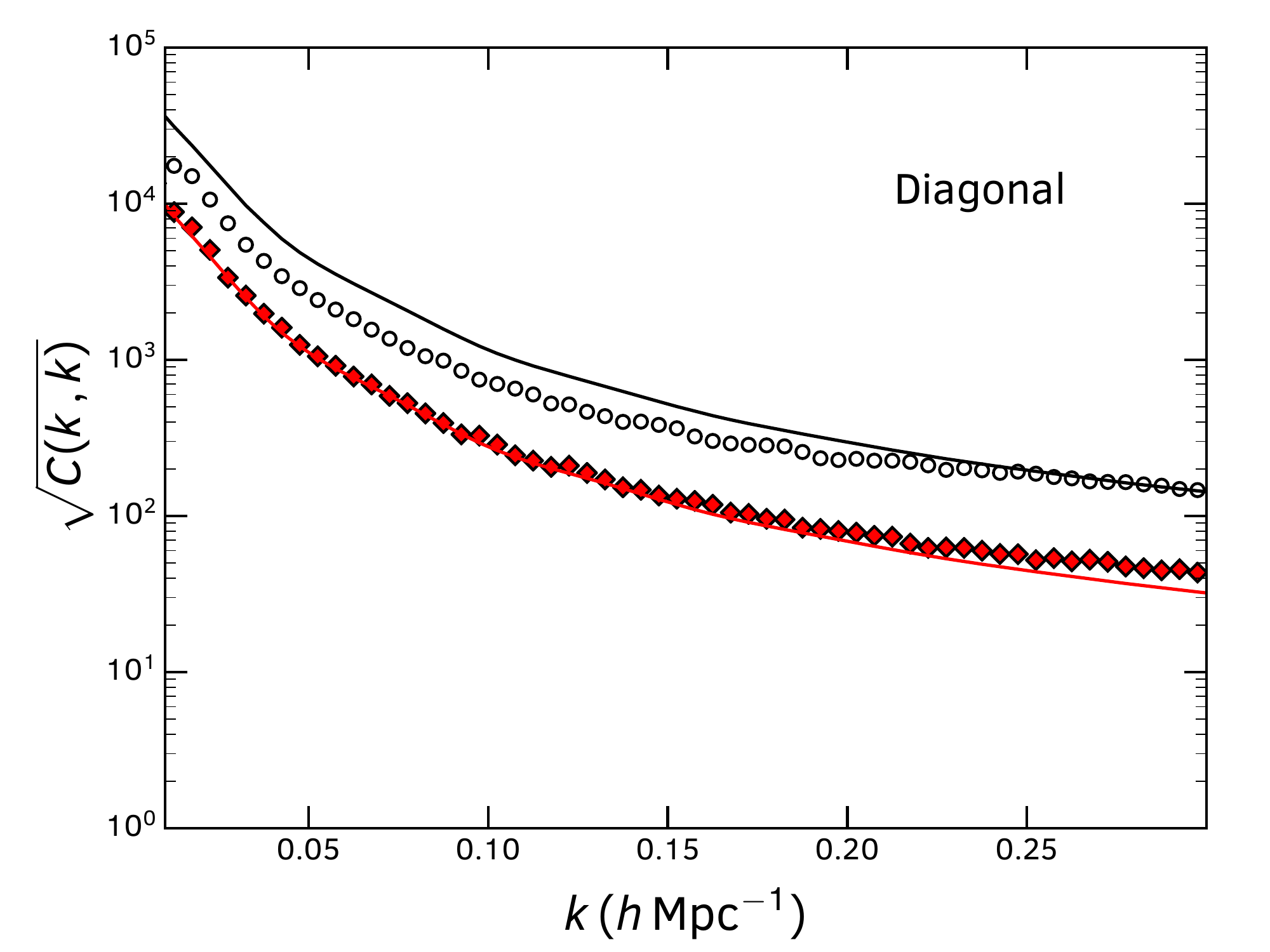}}\\
\subfloat{\includegraphics[width=0.5\textwidth]{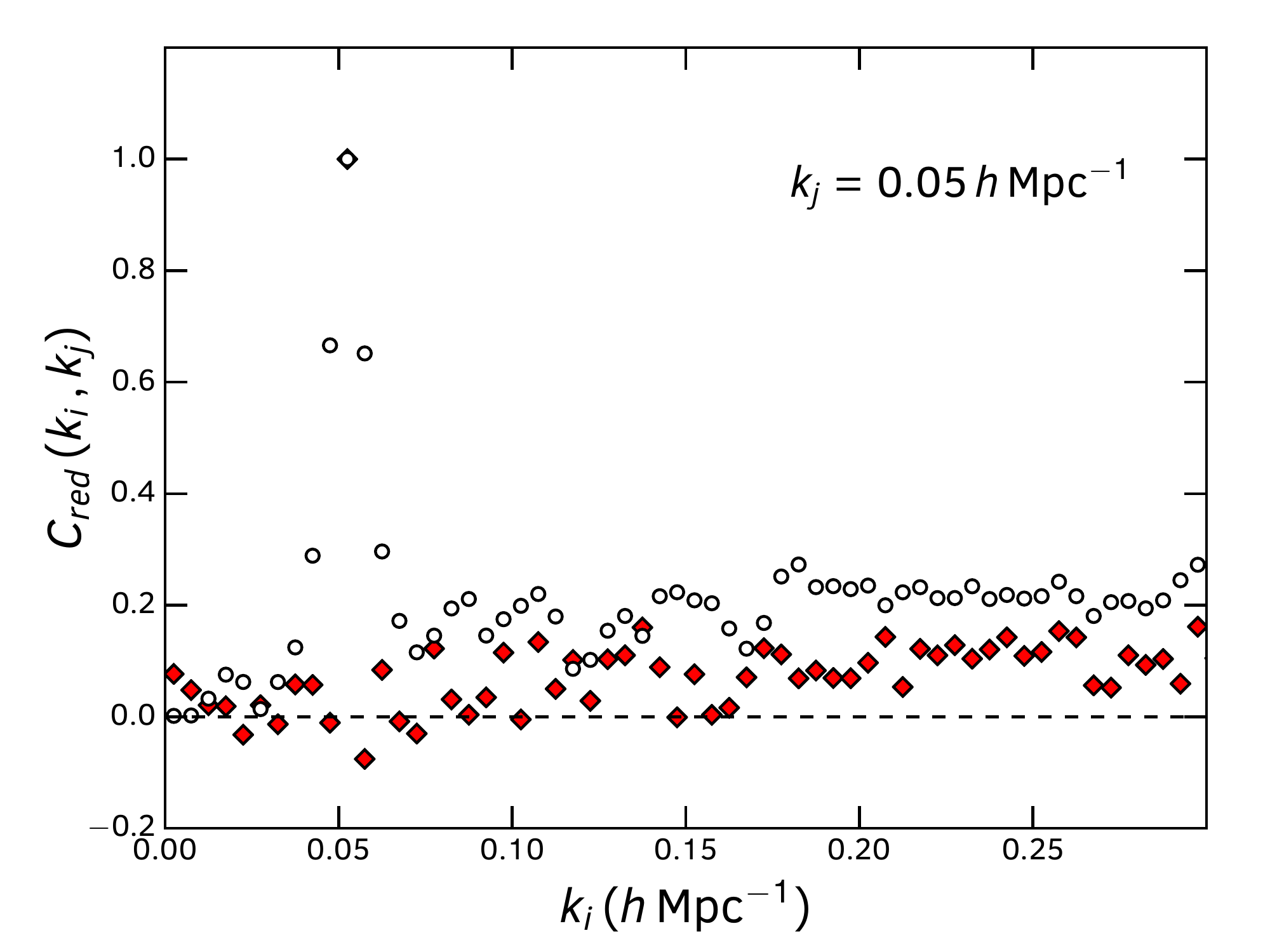}}
\subfloat{\includegraphics[width=0.5\textwidth]{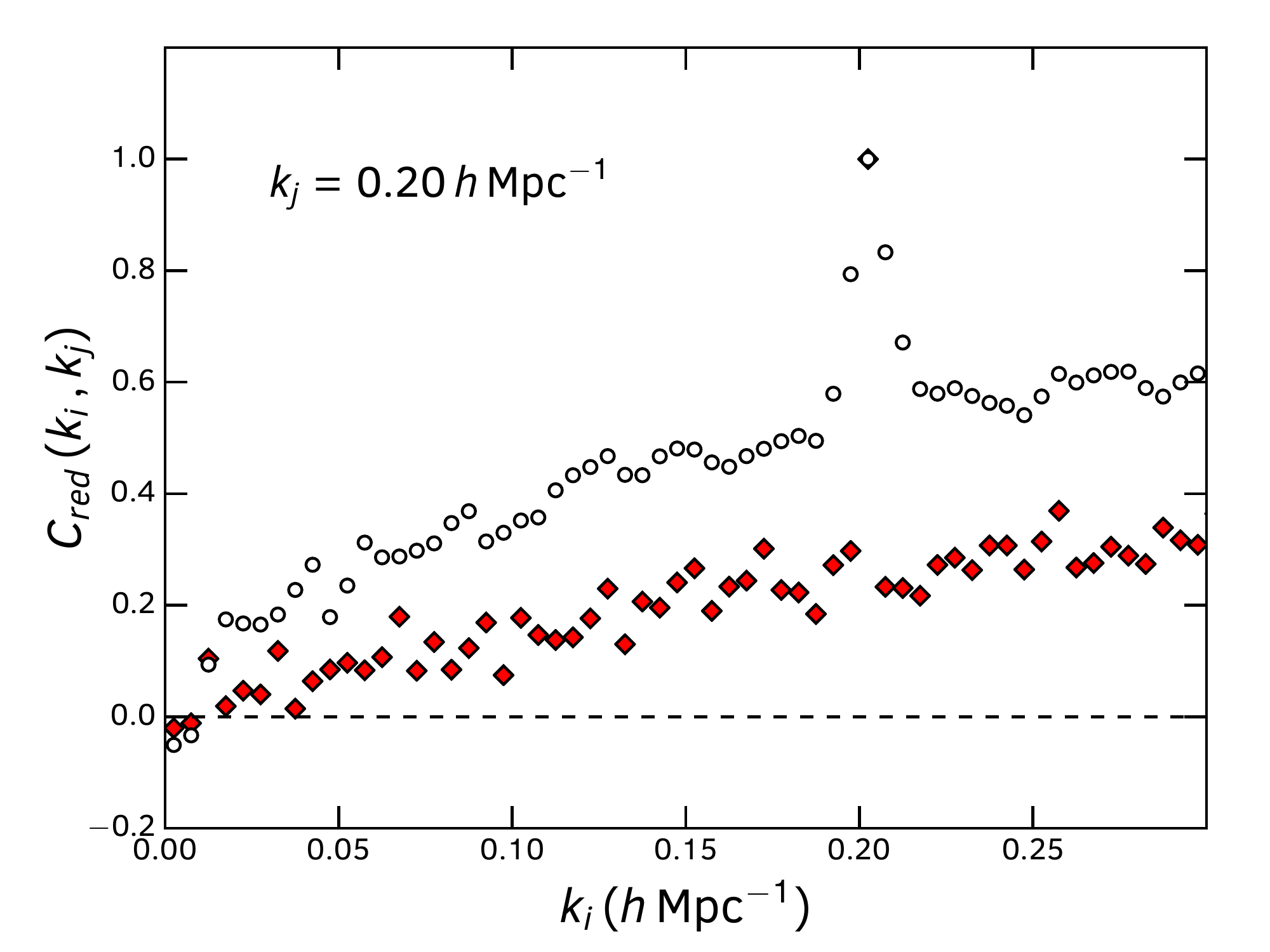}}
\caption{A comparison of the measured covariance matrices for the MGS mocks with and without cutting out the survey window function. Upper left shows the average power spectra of the masked and unmasked simulations. Upper right shows the corresponding covariance matrices measured from the simulations (points) alongside the Gaussian predictions (lines) calculated using the average power spectra and the effective volume of the survey or simulation \citep{Tegmark1997}. The lower two panels show slices through the correlation matrix. The horizontal dashed line is to guide the eye and shows the prediction for a Gaussian field (a $\delta$-function at $k_{i}=k_{j}$).}
  \label{fig:masked_plots}
\end{figure*}

We first look at an example of how the window function changes the covariance matrix. If one has simulations that cover the full survey volume, a brute-force calculation of the covariance matrix including the window function is a simple process; each mock catalogue is masked and subsampled to reproduce the angular and radial distribution of the observed galaxy field. Fig.~\ref{fig:masked_plots} shows the results of this process on the power spectrum and covariance matrix using a set of 500 mock galaxy catalogues originally created for analysis of the Sloan Digital Sky Survey Main Galaxy Sample \citep{Ross2015}. The construction of these mock catalogues is presented in \cite{Howlett2015a}.

From Fig. ~\ref{fig:masked_plots}, we can first identify the familiar way in which the window function reduces the measured power spectrum on large scales, both due to the ``integral constraint'', where modes larger than the survey are not captured, and due to the correlation of large and small scale modes. On top of this, there are several ways in which the window function affects the covariance matrix. Firstly, the amplitude of the covariance is modified by the change in volume between the large cubic simulations and the masked mock catalogues. As the survey volume is smaller than the volume of the simulations this is seen as an increase in the overall amplitude of the covariance matrix. 

However, the effect of the window function if more than just a simple volume scaling, otherwise the large scale measurements from the masked mocks would match the Gaussian prediction for the MGS survey effective volume. The window function correlates different modes in the power spectrum, which is the same as reducing the diagonal elements of the covariance matrix below the Gaussian prediction and increasing the off-diagonal elements. This is seen in the slices of the correlation matrix in Fig.~\ref{fig:masked_plots}, where the peak at $k_{i}=k_{j}$ is significantly broadened in the masked mocks. The non-zero values in the correlation matrix for $k_{i}$ far from $k_{j}$ are seen in both the cubic and masked simulations and arise not due to the window function but due to contributions from higher order clustering and shot-noise terms in the covariance matrix (See Appendix~\ref{sec:appcov}). The change in the amplitude of these terms between the cubic and masked mocks arises due to the relative decrease in diagonal covariance for the masked mocks.

Although time consuming, the convolved power spectrum covariance matrix can be derived and written down analytically \citep{Smith2016}. The full expression and some key steps towards it's derivation are given in Appendix~\ref{sec:appcov}. Although this analytic expression exists, actually calculating the convolution between the window function and the two-, three- and four-point clustering terms requires integrating over 3 $\bk$-vectors (9 integrations in total). Additionally, as the power spectrum and covariance matrix is typically measured in bins, these integrals must be performed for each $\bk$-vector in each bin of interest. Even if the power spectrum, bispectrum and trispectrum could be modelled perfectly, this complexity makes a full theoretical calculation of the convolved covariance matrix practically impossible. Nonetheless, as we will show in the next section, the effects of the window function can still be well modelled analytically for most current and future surveys under some assumptions.

\subsection{Analytic window function convolution} \label{sec:analytic}

To develop our analytic approach, we begin with the assumption that the convolution of the covariance matrix with the window function only occurs on large scales where the covariance matrix is approximately Gaussian, and on smaller scales, where higher order clustering terms become important, the convolution is negligible. Whether or not this assumption is valid will depend on the exact window function of the survey, and it may not hold for very small volume surveys, or narrow pencil-beam surveys. For such surveys, the window function may not tend to zero as rapidly as we go to small scales, and we would be required to consider the convolution with the higher order clustering terms shown Eq.~\ref{eq:appfinal}. However, we will show that it works well for the MGS galaxy sample, which has a small cosmological volume even compared to other surveys of its generation. As such, we expect this method to work very well for larger next generation surveys.

Our first assumption is equivalent to only calculating the first term in Eq.~\ref{eq:appfinal}. If we now assume that the power spectrum is a constant value $P$ over the coherence length of the window, \citet{Feldman1994} showed (FKP; their equation~2.4.6) that this can be written as
\begin{equation}
  \mathsf{C}_{{\rm cst} P}(P(k)) = \frac{2}{N_iN_j}\sum_{i,j}\left|PG_{2,2}(\bk_{i} - \bk_{j})+G_{1,2}(\bk_{i}-\bk_{j})\right|^2,
  \label{eq:Cov_cstP}
\end{equation}
for shells $i$ and $j$ with $N_i$ and $N_j$ modes ${\bf k}_i$ and ${\bf k}_j$ in each. The modes ${\bf k}_i$ are constrained to lie in the shell such that $k < |\bk_{i}| < k+\delta k$ and similarly for $\bk_{j}$ and we have simply renamed the first term in Eq.~\ref{eq:appfinal} under these conditions $\mathsf{C}_{{\rm cst} P}(P(k))$ to emphasise that the power spectrum is assumed constant. A more rigorous derivation of this, given the full equation in Eq.~\ref{eq:appfinal}, can be found in \cite{Smith2016}. We have defined $G_{\ell,m}(\bk)$ as in Eq.~\ref{eq:appgterms}. Qualitatively, the first $G$-term in Eq.~\ref{eq:Cov_cstP} is the normalised Fourier transform of the weighted density field, whilst the second is the shot noise component. For $i=j$, Eq.~\ref{eq:Cov_cstP} is valid where the power is constant across the bin, rather than across the coherence length. 

In order to extend Eq.~\ref{eq:Cov_cstP} to include cross-correlations between bins $i \ne j$, with different power spectrum amplitudes (but constant within each bin), we develop a method to account for the relative impact on the covariance from power leaking from bin $i$ into $j$ and separately from bin $j$ into bin $i$. This is based on the idea that the window function introduces additional covariance between bins, but does not change the amount of information on a mode-by-mode basis. Our ansatz is that the Gaussian part of the covariance matrix under the influence of the window function, which we denote $\bC^{W}$ can then be written as
\begin{equation}
  \mathsf{C}^{W}(k_{i},k_{j}) = \frac{\mathsf{C}_{{\rm cst} P}(P(k_{i}))N_i+\mathsf{C}_{{\rm cst} P}(P(k_{j}))N_j}{N_i+N_j}.
  \label{eq:Cov_cstPij}
\end{equation}
Here we have considered that the window ``spreads'' power $P(k_{i})$ from the $N_i$ modes in bin $i$ into bin $j$, and the power $P(k_{j})$ from the $N_j$ modes in bin $j$ into bin $i$, giving rise to covariances caused by both. 

We implement our approach using a synthetic random catalogue as in the standard FKP method of estimating the power spectrum, replacing the integrals over volume required to calculate the $G$-terms in Eq.~\ref{eq:Cov_cstP} with sums over points randomly placed within the survey mask. The convolved power spectrum within each bin is used as the constant value of $P$, however this is easily computed too by convolving the power spectrum measured in the small volume mocks with the window function analytically \citep{Percival2001,Ross2013}. The following steps are required to calculate $\bC_{{\rm cst} P}$ for a given bin $i$,
\begin{enumerate}
\item{Assign the correct function of number density, weights and convolved power spectra for each random point to a grid in real-space in order to calculate the $G$-terms.}
\item{Fourier transform this grid and calculate the normalised squared modulus value for each $\bk$ as required in Eq.~\ref{eq:Cov_cstP}.}
\item{Now set up the power spectrum squared for the current bin on the grid. Because the power spectrum is assumed constant, we can defer including the amplitude of the convolved power till we evaluate Eq.~\ref{eq:Cov_cstPij}. The power on the grid is then simply one if the $\bk$-vector corresponding to each gridcell is in the current bin, and zero otherwise.}
\item{The power spectrum and $G$-terms on the grid must now be convolved. This is done by inverse Fourier transforming the two grids and multiplying them together.}
\item{Finally perform the sum over all gridcells belonging to the bin, simultaneously counting how many modes are in that bin.}
\end{enumerate}

Once we have computed Eq.~\ref{eq:Cov_cstP} and the number of modes in each bin, it is a simple exercise the evaluate the analytic, convolved covariance matrix. We will show the effectiveness of our method in the following sections. 

\begin{figure*}
\centering
\subfloat{\includegraphics[width=0.5\textwidth]{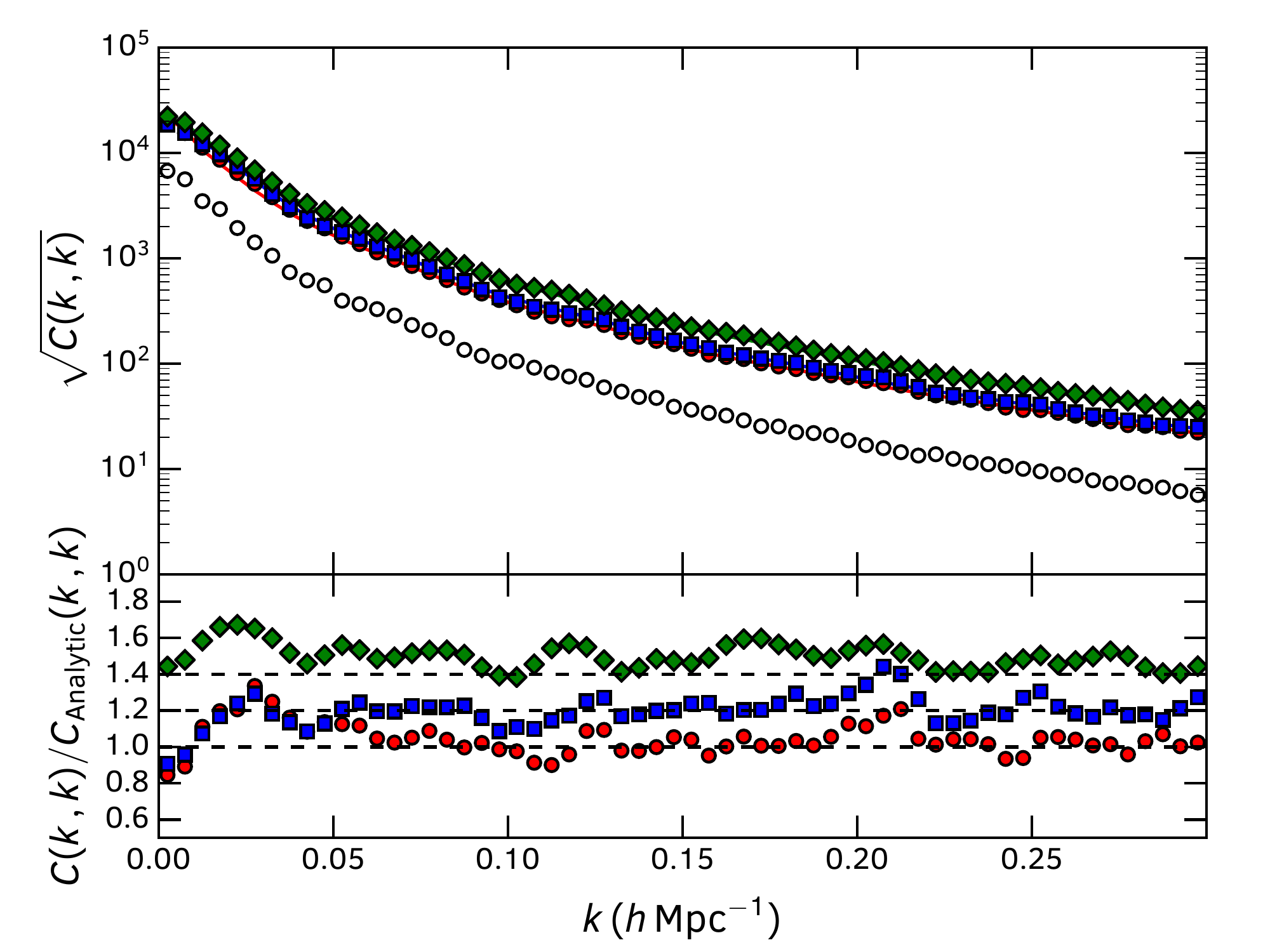}}
\subfloat{\includegraphics[width=0.5\textwidth]{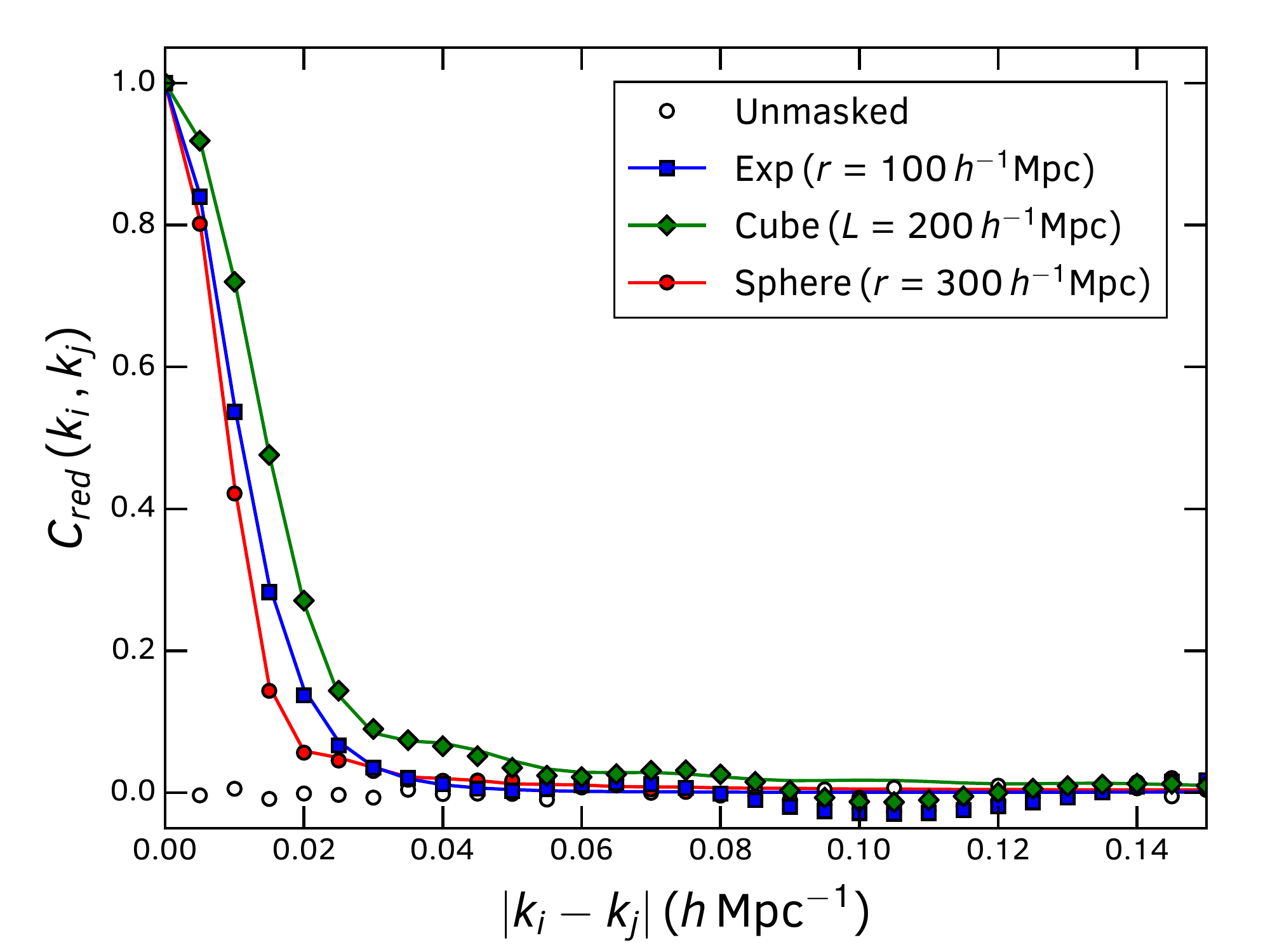}} 
\caption{A comparison of the measured and analytic covariance matrices for 500 Gaussian Random Fields. Left shows the diagonal elements of the covariance matrix, with the lower panel showing the ratio between the measurements and theory, offset by 0.2 (0.4) for clarity. Right shows the off-diagonal elements of the correlation matrix. Points correspond to measurements, whilst lines are the analytic calculation. Different symbols/colors show the different window functions: open circles are the measurements from the unmasked mocks; red circles/lines are for a spherical tophat with $r=300\mpcoh$; blue squares/lines are for an exponential weighting with scale length $r=100\mpcoh$; and green diamonds/lines are for a cube of edge length $L=200\mpcoh$. The horizontal lines in the lower left panel are to guide the eye; we expect the ratio between the measurements and theory to be 1 (modulo the offset we have included).}
  \label{fig:gausstest}
\end{figure*}

\subsection{Tests on GRF's} \label{sec:grf-test}

We begin by testing our method for including the window function analytically on the same Gaussian random fields used in Section~\ref{sec:motivation}. We take the 500 GRF's of size $L=1280\mpcoh$ and calculate the covariance matrix under a set of simple window functions. For our tests we use a spherical tophat of radius $300\mpcoh$, a cube of edge length $200\mpcoh$ taken from the middle of the simulation, and an exponential weighting function with scale length $100\mpcoh$. For these three cases we calculate the covariance matrix using the brute force method and using a random catalogue and the method in Section~\ref{sec:analytic}. All of the convolutions for our analytic calculation are performed on a grid of the same size as was used to generate the GRFs and no weighting is applied to the density field other than that of the window function. The measurements and theory are shown in Fig.~\ref{fig:gausstest}.

As the Gaussian random fields have no bispectrum or trispectrum components, and no shot-noise, we expect our analytic covariance matrix to agree very well with the measurements. Our method should only break down where the window function is small enough or complex enough that the power spectrum is no longer constant across its coherence length. For the three different window function we test, we find excellent agreement between the measured covariance matrix and our theory. The change in amplitude of the covariance matrix due to the inclusion of a window function is well recovered, which can be seen comparing the diagonal elements of the covariance matrices, as is the introduction of off-diagonal covariance due to the convolution with the window function. The largest discrepancy between the two is seen in the diagonal elements for the cubic window, however this window is quite an extreme case and results in a strong suppression of large scale power due to the small volume of the `survey'.

\subsection{Tests on a realistic survey}  \label{sec:mock-test}

\begin{figure*}
\centering
\subfloat{\includegraphics[width=0.5\textwidth]{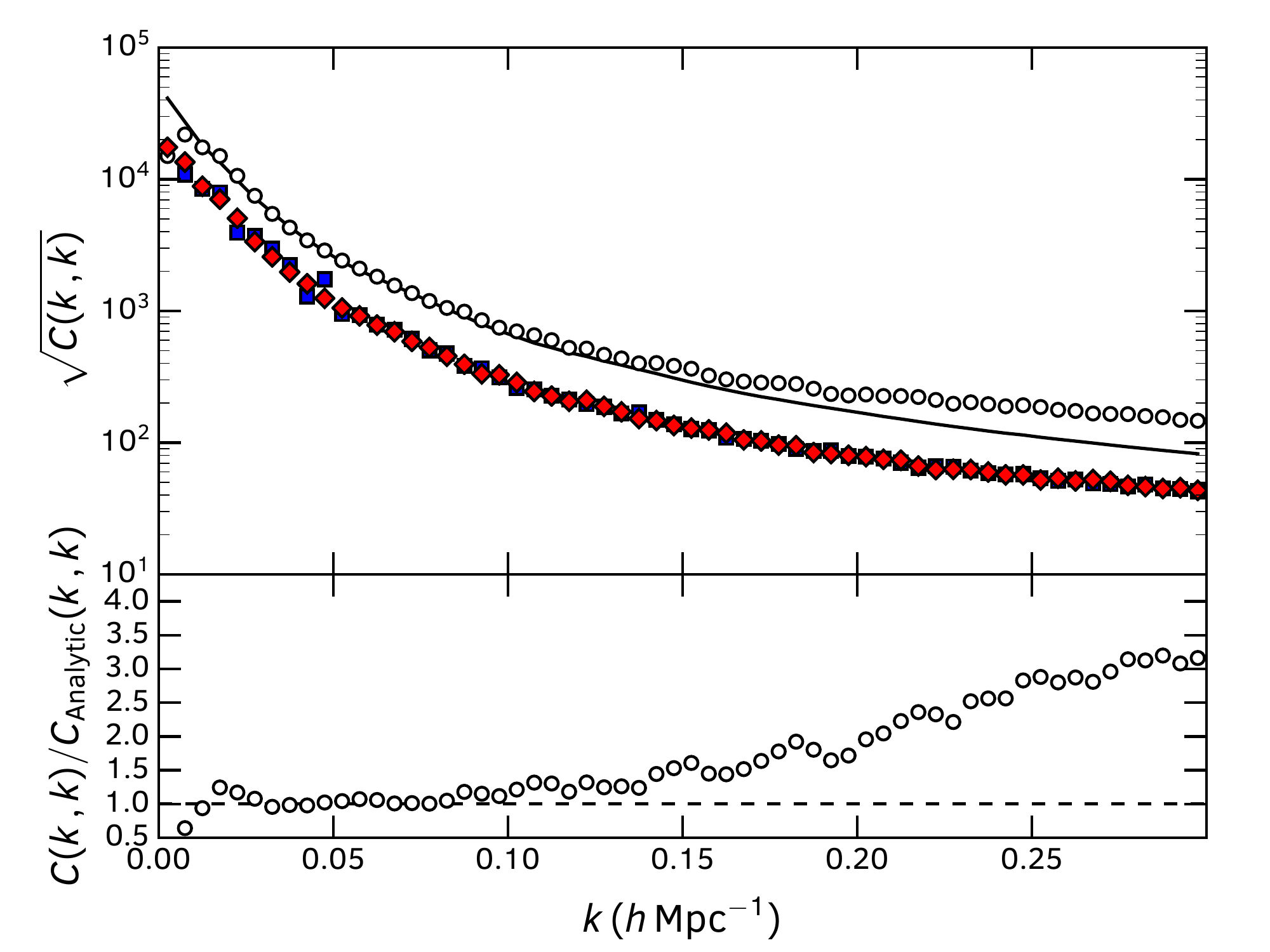}}
\subfloat{\includegraphics[width=0.5\textwidth]{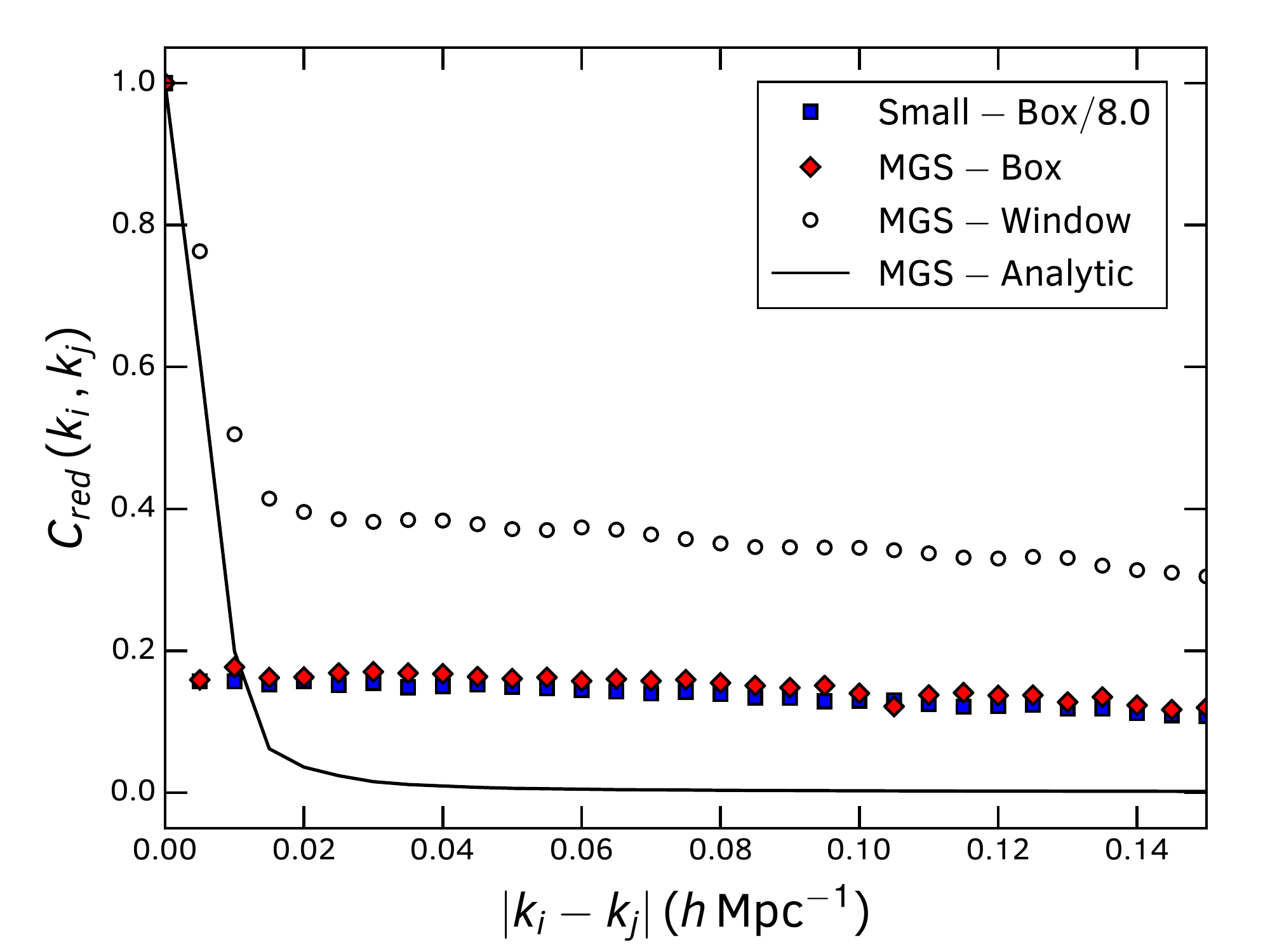}} 
\caption{A comparison of the measured and analytic covariance matrices for the 500 masked MGS mock galaxy catalogues, plotted in the same way as Fig.~\ref{fig:gausstest}. Open points show the measurements from the masked mocks catalogues whilst the solid line is the analytic prediction. The red diamonds and blue squares show the covariance matrices of the 500 large and 4000 small volume cubic MGS simulations, with the latter divided by a factor of 8 (the ratio of the volumes) to show the consistency between these two sets. We see that the analytic prediction matches the diagonal elements of the measured covariance well on large scales, but diverges on non-linear scales and for the off-diagonal components due to the absence of higher order clustering and shot-noise terms.}
  \label{fig:MGStest}
\end{figure*}

In order to test the application of our method to an actual galaxy sample, including the effects of shot-noise, higher order clustering and weights, we next match to the MGS sample introduced in Section~\ref{sec:window}. Our analytic calculation requires only a random catalogue and an estimate of convolved power spectrum. Because this same random catalogue is used to estimate the power spectrum from the data it is trivial to include the effects of weighting (both systematic and optimal FKP weights) as long as these have been given to each of the random points. For all of our measurements and calculations we assume a FKP weighting with fixed power spectrum $\bar{P}=10000\,h^{-3}{\rm Mpc}^3$, and work on a $512^{3}$ grid of side $1280\,h^{-1}{\rm Mpc}$.

For our input power spectrum, and for later use, we generate a suite of 4000 \textsc{l-picola} simulations each 1/8th the size of the original MGS mocks, but with the same mass resolution. We identify halos in each of our dark matter fields and populate them with galaxies using the same procedure and HOD model as was used for the original MGS mock catalogue sample in \cite{Howlett2015a}. The average clustering of these simulations matches the original mock catalogues exactly, except on large scales where there is insufficient volume to measure the power spectrum accurately in the small volume simulations. Because of the procedure we have used to simulate these galaxy mocks, we expect both the large and small volume sets to reproduce the clustering in the data equally well. Using small volume simulations for the clustering also means our method automatically incorporates the effects of shot noise and galaxy bias in a more accurate way than if we had attempted to model this theoretically. Finally, the time taken to produce the 4000 small volume mock catalogues is actually less than that required to produce the set of 500 larger mocks because of the imperfect scaling of the codes used for our simulations. We would expect nearly all codes used for such simulations to behave similarly.

Using the 4000 small volume mocks and the random catalogue, we calculate the analytic covariance matrix and compare it to the brute force covariance matrix measured from the 500 large volume mock catalogues in Fig.~\ref{fig:MGStest}. On large scales the analytic perscription recovers the effects of the window function on the diagonal covariance matrix extremely well, matching both the change in the overall amplitude of the covariance due to the change in effective volume, and the relative re-weighting of large scale diagonal and off-diagonal covariance due to convolution with the window function. However, as our method only solves the Gaussian part of the covariance matrix, it under-predicts the diagonal covariance on small scales and the off-diagonal covariance for $k_{i}$ significantly different from $k_{j}$. The origin of this additional covariance is not the window function, but rather higher order clustering and shot-noise, in particular the trispectrum term. In the following section we will demonstrate how to include these final components using the small volume mocks.

\section{Combining small volume mocks and our analytic methods} \label{sec:endresult}
Thus far we have advocated that the computational burden of generating sizeable numbers of mock galaxy catalogues for covariance matrix estimation can be reduced by `scaling' the covariance matrix measured from a set of small volume simulations that do not necessarily fit the full survey volume. Alternatively this allows one to improve the accuracy of the covariance matrix estimation for a fixed computational time. We have also presented methods to analytically correct for the effects of modes outside the simulation and the survey window. In this section, we bring everything together and show how these methods can be combined with the small-volume covariance matrix to fully recover that measured from a set of full-size simulations. In doing so we will correct the final discrepancies between the analytic method and the brute force estimation highlighted in Fig.~\ref{fig:MGStest} and the previous section.

When calculating the analytic window function we assumed that only the Gaussian part of the covariance matrix is convolved by the survey window. Considering this, we can approximate the binned covariance matrix measured from a set of masked mocks as the non-Gaussian parts of the covariance matrix measured from the small-volume mocks, multiplied by a factor based on the ratio of the effective volumes between the survey and the simulation, plus the convolved Gaussian part of the covariance matrix, which we have shown can be calculated analytically. A mathematical derivation of this is presented in Appendix~\ref{sec:appcov2}. 

In deriving our method we introduce an additional approximation on top of the assumption that only the Gaussian part of the covariance matrix is convolved with the window function, namely that each `group' of higher order terms (trispectrum, bispectrum, power spectrum and constant) in the covariance matrix scales with the same effective-volume-based factor. This is only true for a survey with constant number density, otherwise the different terms have different shot-noise dependencies and hence slightly different scaling factors. An alternative description is given in Appendix~\ref{sec:appcov2}, where we associate our approximation with an additional \textit{residual} component of the covariance matrix that depends on the bispectrum and power spectrum.

In practice, we find that applying this approximation and ignoring the \textit{residual} covariance, gives very reasonable results for the diagonal and off-diagonal covariance, as the dependence of the bispectrum and power spectrum terms on the shot-noise means they only become important on highly non-linear scales. If necessary, given a model/measurement for the bispectrum and power spectrum, these additional terms could be included more accurately quite easily, as shown in Appendix~\ref{sec:appcov2}.

\begin{figure*}
\centering
\subfloat{\includegraphics[width=0.5\textwidth]{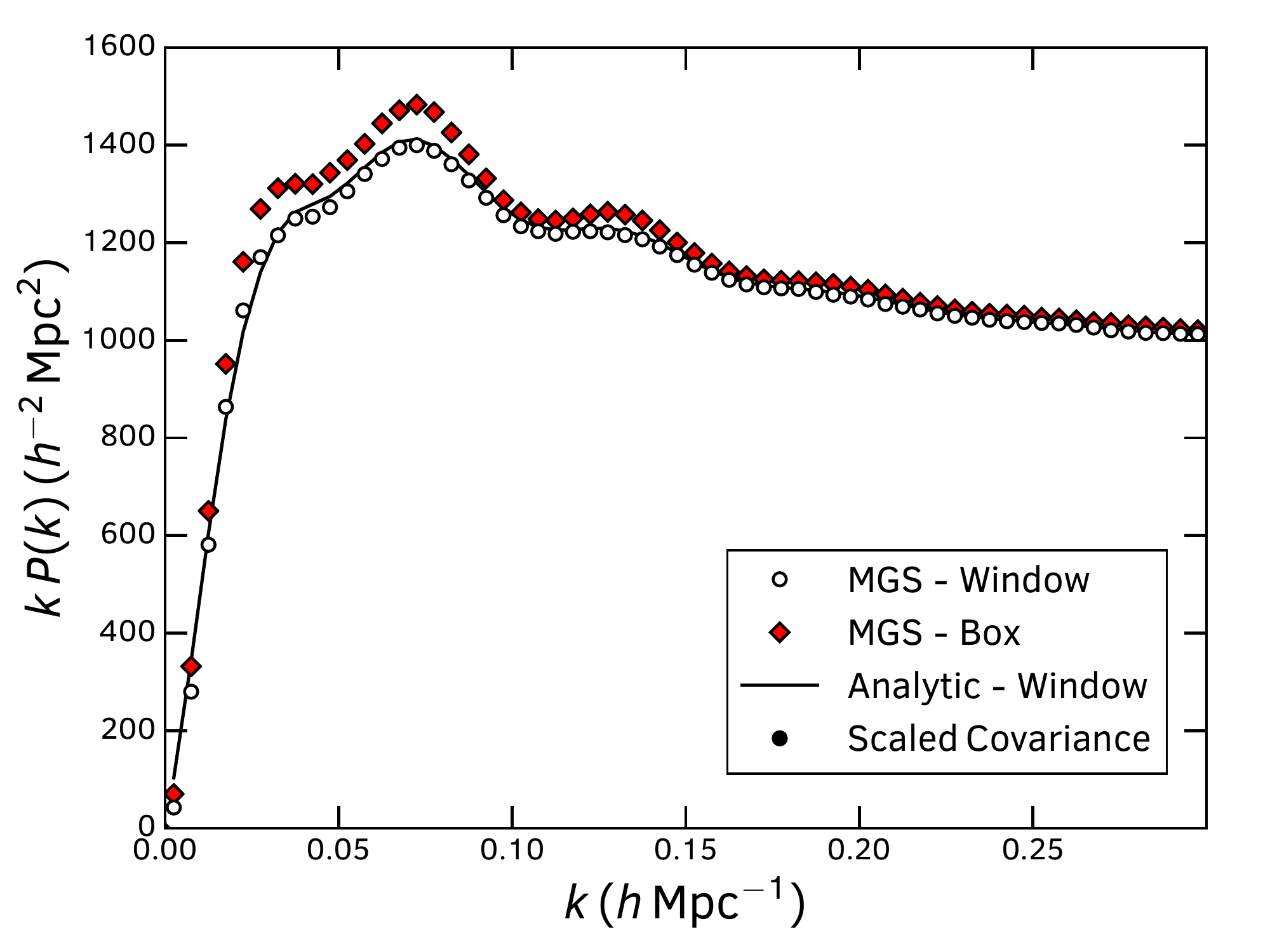}}
\subfloat{\includegraphics[width=0.5\textwidth]{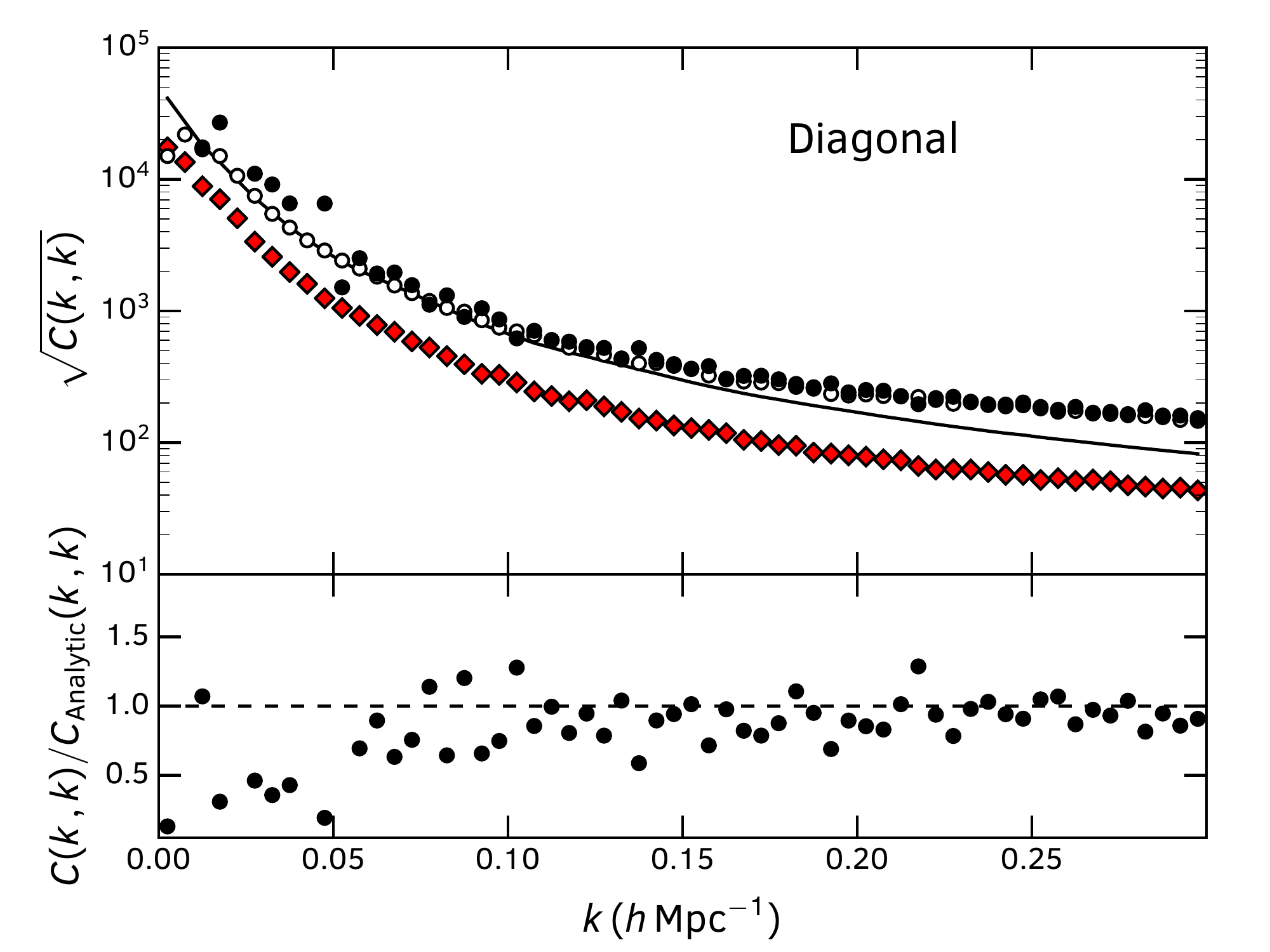}}\\
\subfloat{\includegraphics[width=0.5\textwidth]{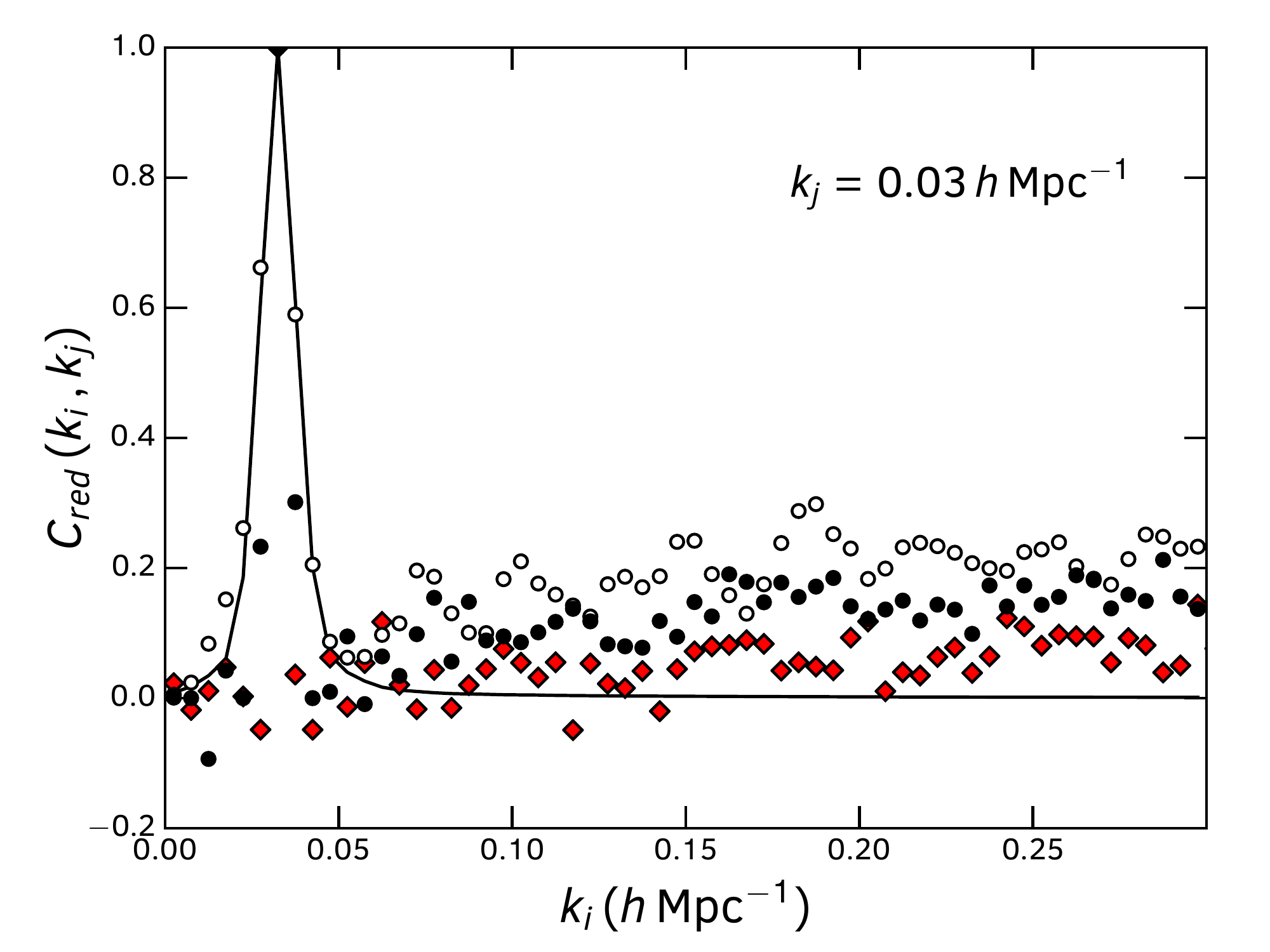}}
\subfloat{\includegraphics[width=0.5\textwidth]{Figure8_middle_left.pdf}} \\
\subfloat{\includegraphics[width=0.5\textwidth]{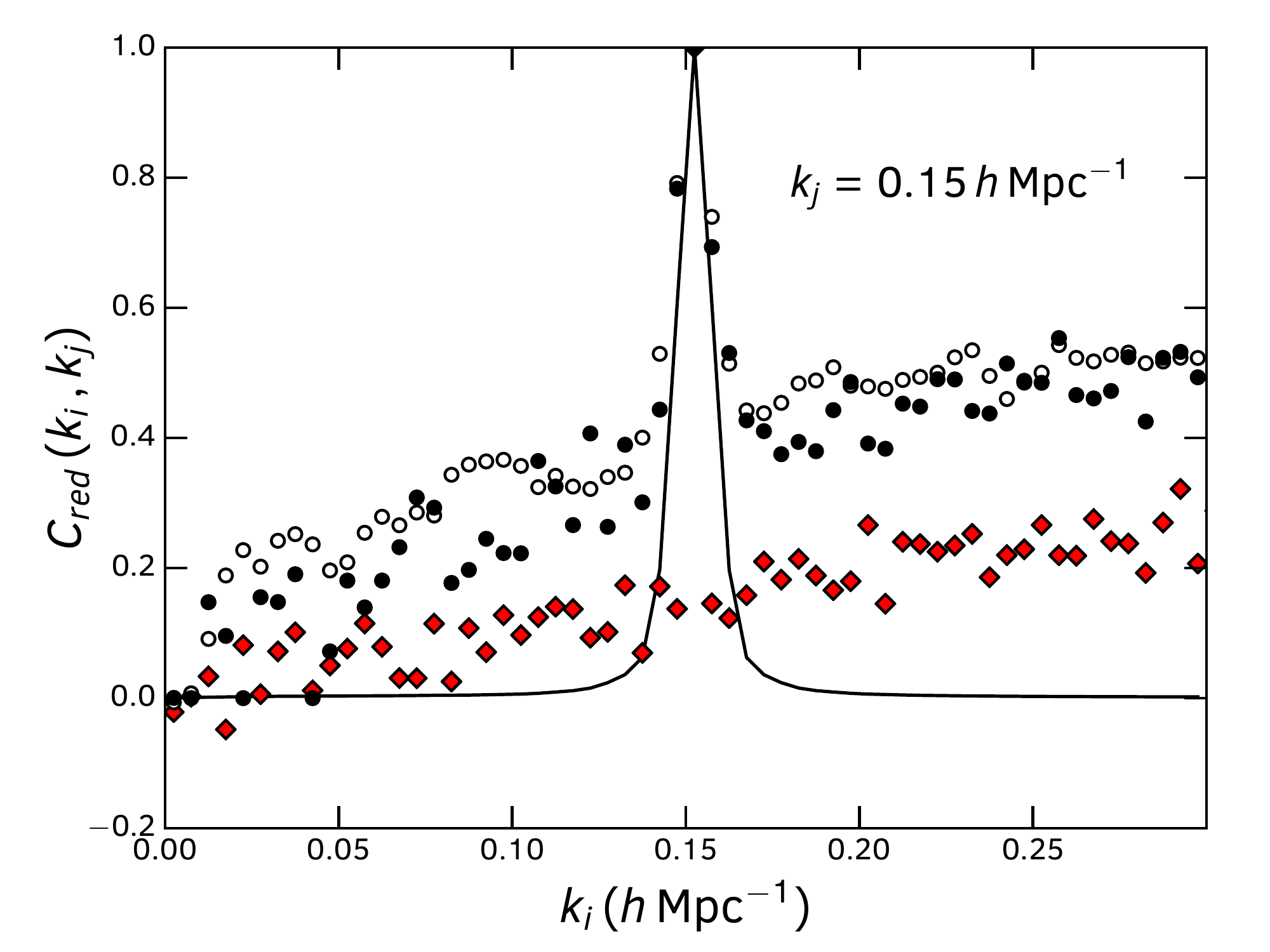}}  
\subfloat{\includegraphics[width=0.5\textwidth]{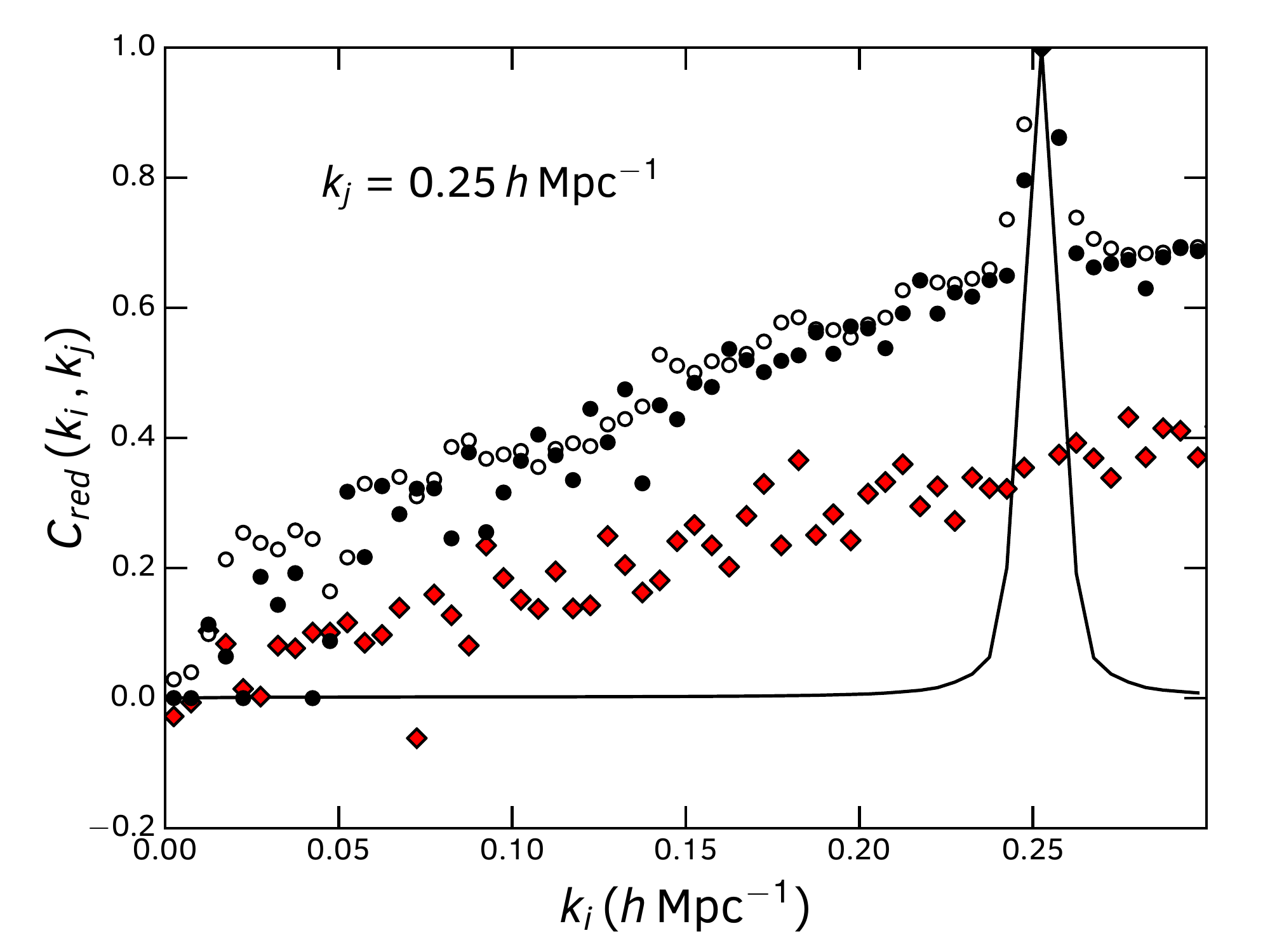}}    
\caption{A comparison of the measured and analytic covariance matrices for the MGS galaxy mocks. The top left and right panels show the power spectra and the diagonal of the covariance matrix, whilst the remaining panels are slices through the correlation matrix, highlighting the off-diagonal terms. The lower plot in the top right panel is the ratio of the measured covariance matrix against our analytic, scaled covariance. In all panels, red diamonds show measurements from the 500 large volume cubic simulations, open points show the measurements from the masked simulations which include implicit convolution with the survey mask. By combining the analytic window function convolution (black line) with a rescaled version of the small volume covariance matrix and the Gaussian prediction for the small volume we obtain our scaled covariance matrix (black points) which recovers the behaviour of masked simulations remarkably well on both large and small scales. This proves that our rescaling method is a viable method of reproducing the covariance matrix and large volume simulations covering the full survey volume are not required.}
  \label{fig:scaled_covariance_MGS_plots}
\end{figure*}

In Fig.~\ref{fig:scaled_covariance_MGS_plots}, we show the final result of the procedure to convert the small volume cubic covariance matrix into that measured from a set of masked and subsampled mocks catalogues, the key outcome of this work. Again, the necessary $G$-terms are computed using a summation over the random catalogue. We plot the measured, masked covariance for the MGS mocks, but this time against our full method combining both the analytic window function convolution and the cubic covariance matrix. We find excellent agreement between the diagonal and off-diagonal components of the true covariance matrix and the matrix estimated with our new method on all scales of interest to large scale structure surveys. By including the higher order clustering and shot-noise terms from the small-volume cubic simulations, we have corrected the discrepancies seen in Fig.~\ref{fig:MGStest}.

\subsection{Extending to redshift space}

Throughout this work we have dealt only with the covariance matrix of the real-space spherically averaged power spectrum. Redshift Space Distortions give rise to additional correlations between the anisotropic window function and power spectrum that are not accounted for completely by the above method. In the case of no window function, the bin-averaged covariance matrix can still be computed and the scaling we advocate is still applicable, but when we are bin-averaging in the presence of the window function our analytic window function calculation based on FKP no longer fully describes how the power leaks from one bin to another.

Eqs.~\ref{eq:Cov_cstP} and \ref{eq:Cov_cstPij} require the power spectrum to be constant within the bin being considered. In general, this is not the case for binned Legendre moments of the redshift-space, line-of-sight dependent, power spectrum. Redshift space distortions, together with changes in the line-of-sight across a survey, mean that the clustering depends on spatial position across a survey and the simple window function convolution derived in equation~2.1.6 of FKP breaks down. \cite{Wilson2017} show that the measured line-of-sight dependent power spectrum moments \citep{Bianchi2015,Scoccimarro2015} can be described by a sum of convolutions of the plane-parallel moments with different windows. Extending their derivation to determine the covariance as well as the convolved power would theoretically be possible, but it would lead to a complicated expression with a large number of terms for the covariance even for a single mode. 

\begin{figure*}
\centering
\subfloat{\includegraphics[width=0.5\textwidth]{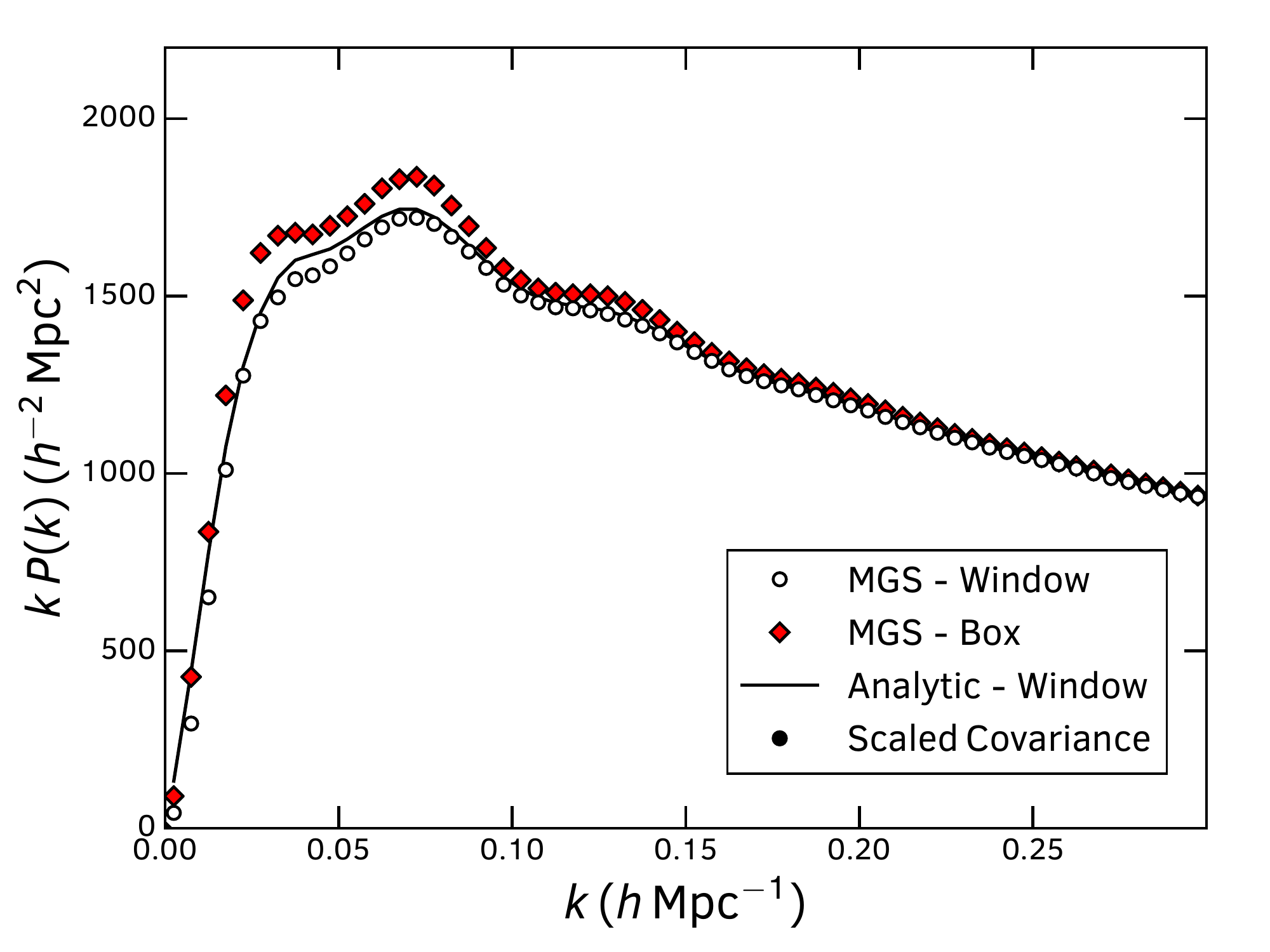}}
\subfloat{\includegraphics[width=0.5\textwidth]{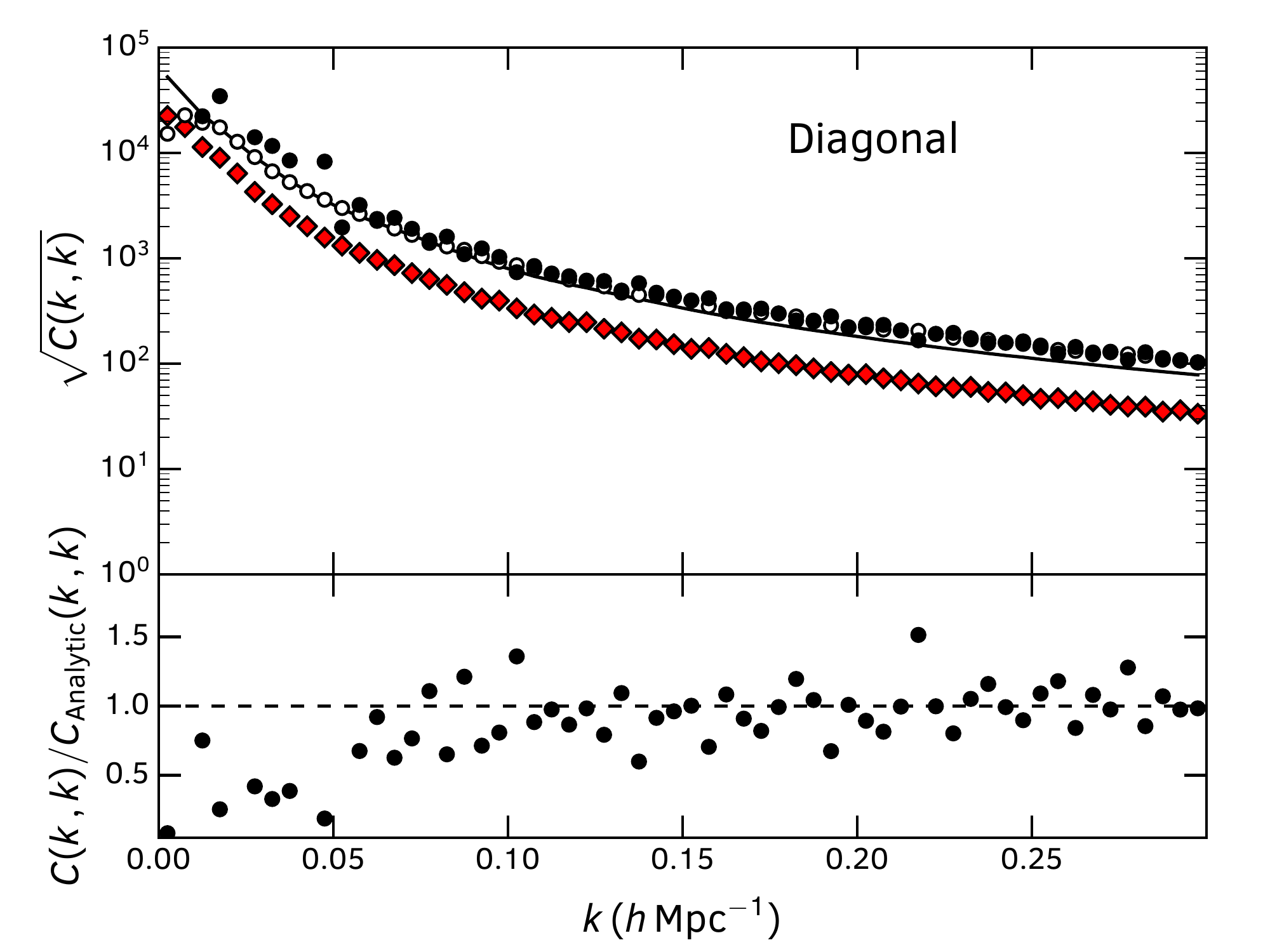}}\\
\subfloat{\includegraphics[width=0.5\textwidth]{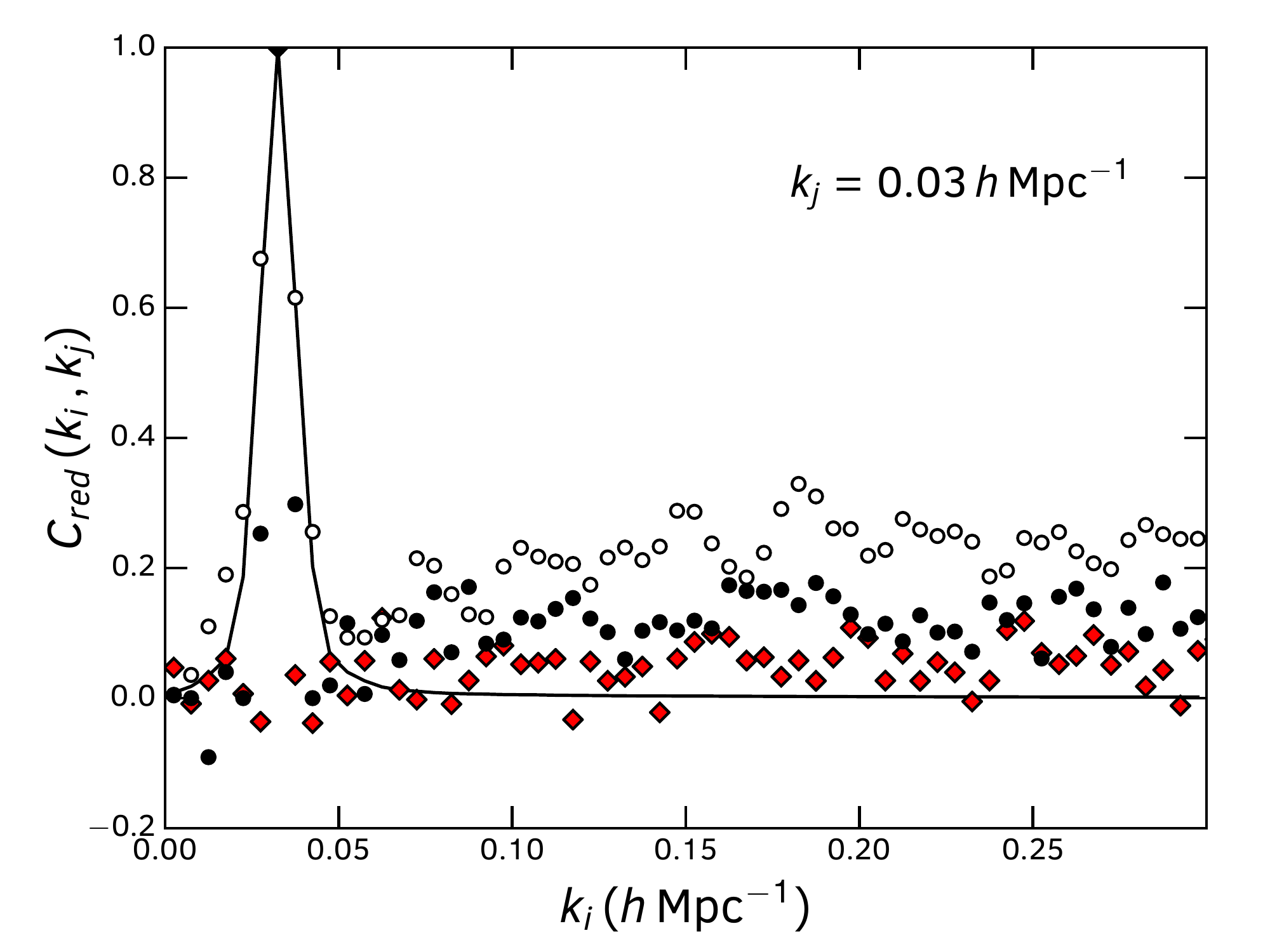}}
\subfloat{\includegraphics[width=0.5\textwidth]{Figure9_middle_left.pdf}} \\
\subfloat{\includegraphics[width=0.5\textwidth]{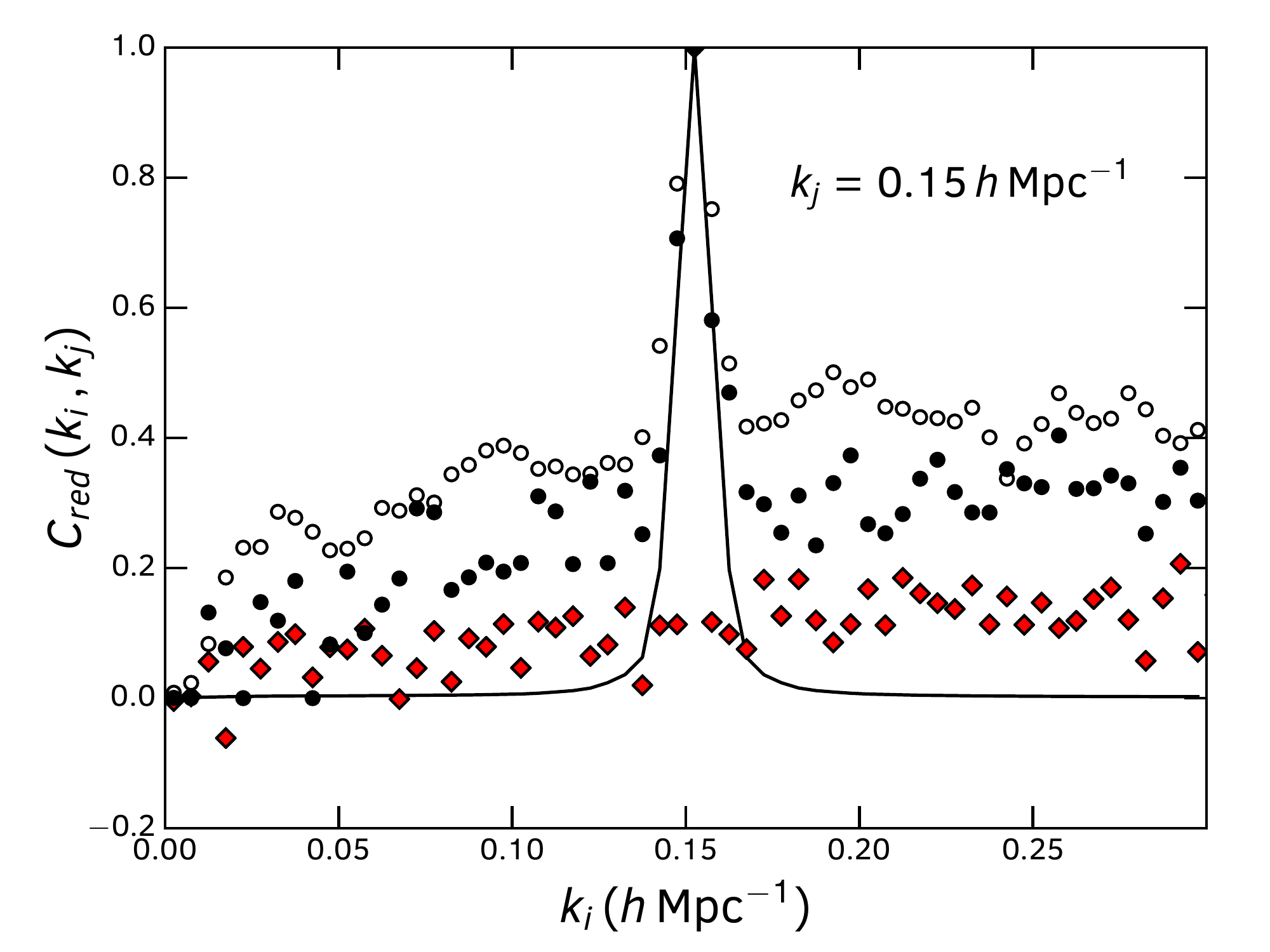}}  
\subfloat{\includegraphics[width=0.5\textwidth]{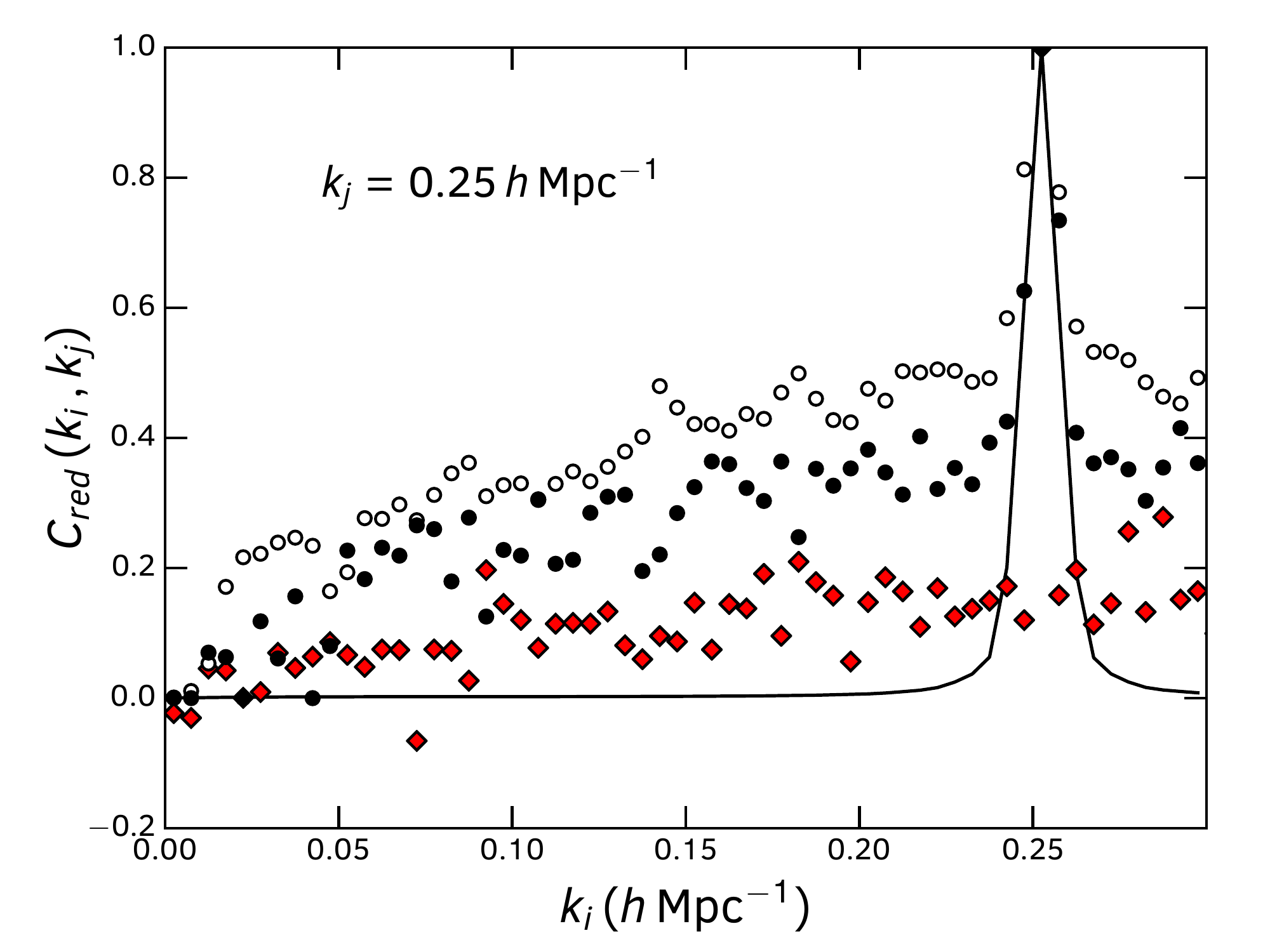}}      
\caption{As for Fig.~\ref{fig:scaled_covariance_MGS_plots} but in redshift space. The masked covariance matrix is now calculated using the redshift space positions of all the mocks catalogues, and we perform our analytic window function calculation using the redshift space power spectrum. Although we do not expect our method to hold for the redshift space power spectrum due to the additional coupling between the power spectrum and window function in redshift space, we still find reasonable results for the MGS sample.} 
  \label{fig:scaled_covariance_MGS_plots_zspace}
\end{figure*}

Regardless of this shortcoming, we find that the coupling between the window and the LOS-dependent power is small for the MGS sample, and that simply replacing the real-space power spectrum with the redshift-space power spectrum monopole in Eqs.~\ref{eq:Cov_cstP} and \ref{eq:Cov_cstPij} provides a reasonable fit to the measured covariance in redshift space. Inaccuracies in modelling the power spectrum and covariance matrix using the spherically averaged window function (which can be seen in the model power spectrum in Fig.~\ref{fig:scaled_covariance_MGS_plots_zspace}) cause discrepancies in the large scale diagonal covariance matrix and a small underestimation of the off-diagonal terms. A more rigorous analysis of this, and application to the higher order multipoles of the power spectrum and their covariance is left for future work.

\subsection{Configuration space}
The combination of analytic methods and measured covariance matrix that we have successfully used to generate the binned covariance matrix of the masked galaxy power spectrum could also be applied to configuration space statistics. As with the power spectrum covariance matrix, the covariance matrix of the correlation function is expected to scale with the volume and number density of tracers; increasing either the volume in which to measure the clustering or the number of galaxies will decrease the amplitude of the covariance matrix. The four-point nature of the covariance matrix demonstrated herein also translates through to configuration space, with the covariance matrix of the two-point correlation function depending on the three- and four-point correlation functions.

The effects of a survey window on the correlation function are easier to model than on the power spectrum as the survey window acts as a multiplicative function rather than a convolution, and the integral constraint can be modelled as a simple additive term. \cite{OConnell2016} present a method to calculate the masked configuration space covariance matrix in the Gaussian regime. The method presented in this work of using small scale simulations to reproduce the effects of the higher order non-Gaussian components as opposed to modelling these analytically is still expected to hold in configuration space, and so could be combined with the formulae from \cite{OConnell2016} to produce the masked covariance matrix of the correlation function. We also leave a broader exploration of this and a derivation of the necessary analytic scaling factor for future work.

\section{Conclusions} \label{sec:conclusion}
In this work we have presented a new method for calculating the covariance matrix of the binned, spherically averaged galaxy power spectrum. Past work on the covariance matrix has focused on methods to reduce the number of realisations required to achieve some numerical accuracy in estimates of the covariance matrix, but these do not address the problem that running even a few hundreds of simulations with the required cosmological volume and resolution for next generation surveys such as DESI and Euclid will be challenging. 

Instead, we have shown that the masked binned covariance matrix can be estimated by combining analytic calculations on large scales, which assumes Gaussianity yet capture the effects of the survey window, with the small scale covariance matrix measured from a set of realistic cubic galaxy mocks that do not have to fit the full volume of the survey. We have also included a method to incorporate the effects of large scale modes that cannot be included in the small volume simulations by using the `Separate Universe' approach, slightly modifying the input cosmology of every simulation within a suite to account for variations in the large scale power.

The benefit of our method is two-fold: Firstly we ease the computational burden of requiring large number of huge, high resolution simulations for covariance matrix estimation. Secondly, we have shown that we can actually improve the error on the covariance matrix in a fixed computational time as the number of simulations we can run will scale approximately with the volume. Analytically scaling the covariance matrix from these simulations conserves the improvement in the error gained by running additional simulations.

As proof-of-concept we are able to reproduce the covariance matrix of the spherically averaged power spectrum measured from a set of 500 full-size masked simulations originally generated for the analysis of the SDSS-II Main Galaxy Sample using only the random catalogue associated with the MGS data and a set of 4000 new simulations each 1/8th the size of the original simulations. The computational time required to generate both ensembles of simulations is approximately the same. The cosmological volume of the MGS is much smaller than for future surveys, and hence the effects of the survey window function of the covariance matrix more severe. The fact that our method does well even in this case means we expect it to perform extremely well for next generation surveys.

However, this work is only the first step towards a viable alternative for covariance matrix estimation. Further effort is required to investigate how to incorporate the effects of Redshift Space Distortions, and test the method in configuration space and on the multipoles of the two-point clustering statistics. Nonetheless this presents an extremely promising route to alleviating one of the greatest computational burdens 
faced by future large scale structure analyses.

\section*{Acknowledgements}
This research was conducted by the Australian Research Council Centre of Excellence for All-sky Astrophysics (CAASTRO), through project number CE110001020.

WJP acknowledges support from the European Research Council through the Darksurvey grant 614030, and from from the UK Science and Technology Facilities Council grant ST/N000668/1 and the UK Space Agency grant ST/N00180X/1. 

Numerical computations were done on the Sciama High Performance Compute (HPC) cluster which is supported by the ICG and the University of
Portsmouth.

This research has made use of NASA's Astrophysics Data System Bibliographic Services and the \texttt{astro-ph} pre-print archive at \url{https://arxiv.org/}. All plots in this paper were made using the {\sc matplotlib} plotting library \citep{Hunter2007}.

\bibliography{/Volumes/Work/ICRAR/LaTeX/massive}{}
\bibliographystyle{mnras}


\appendix
\section{Mathematical expression for the power spectrum covariance matrix} \label{sec:appcov}
FKP give a mathematical derivation for the binned power spectrum measured from a galaxy survey using two point correlations between real and synthetic galaxy catalogues. The same procedure can be used for the covariance matrix of the measured binned power spectrum, although the derivation is much longer and requires one to consider the correlations between four distinct locations.

We first define the power spectrum covariance matrix for a pair of $\bk$-vectors as
\begin{equation}
\mathsf{C}(\bk,\bk') = \langle P(\bk)P(\bk') \rangle - \langle P(\bk) \rangle \langle P(\bk') \rangle,
\label{eq:app1}
\end{equation}
where $P(\bk)$ is the power spectrum for some $\bk$-vector. The binned covariance matrix can then be expressed as
\begin{equation} 
\mathsf{C}(k_{i},k_{j}) = \int_{V_{k_{i}}} \frac{d^{3}k}{V_{k_{i}}} \int_{V_{k_{j}}} \frac{d^{3}k'}{V_{k_{j}}} \mathsf{C}(\bk,\bk').
\label{eq:app2}
\end{equation}
Substituting the power spectrum $P(k)$ in Eq.~\ref{eq:app1} for the FKP estimator and using the fact that the shot-noise is scale-independent, we can write the binned covariance matrix as the correlations between the weighted density field $F(\br)$ at different locations
\begin{equation}
\begin{aligned}
& \mathsf{C}(\bk,\bk') = \int d^{3}r_{1} \int d^{3}r_{2} \int d^{3}r_{3} \int d^{3}r_{4} \,e^{i\bk \cdot (\br_{1}-\br_{3}) + i\bk ' \cdot (\br_{2}-\br_{4})} \\
& \biggl[ \langle F(\br_{1})F(\br_{2})F(\br_{3})F(\br_{4})\rangle - \langle F(\br_{1})F(\br_{3})\rangle \langle F(\br_{2})F(\br_{4})\rangle\biggl],
 \end{aligned}
\label{eq:app3}
\end{equation}
where $F(\br)$ is defined in terms of the number density $\bar{n}(\br)$, and weights $w(\br)$, given to the real and synthetic points at location $\br$, as per Eq.(2.1.3) in FKP. The tedious process of evaluating all the potential correlations between the real and synthetic weighted density fields uses the method given the appendix of FKP and is detailed in \cite{Smith2016}. Substituting all the necessary terms into Eq.~\ref{eq:app3} and Eq.~\ref{eq:app2} in turn gives the final expression for the binned covariance matrix
\begin{widetext}
\begin{align}
\mathsf{C}(k_{i},k_{j}) = & \int_{V_{k_{i}}} \frac{d^{3}k}{V_{k_{i}}} \int_{V_{k_{j}}} \frac{d^{3}k'}{V_{k_{j}}} \biggl\{ \, \biggl|2\int d^{3}q_{1} P(\bq_{1}) G_{1,1}(\bk-\bq_{1})G_{1,1}(\bk '+\bq_{1}) + (1+\alpha)G_{1,2}(\bk+\bk ')\biggl|^{2} \notag \\
+ & \int d^{3}q_{1} \int d^{3}q_{2} \int d^{3}q_{3} T(\bq_{1},\bq_{2},\bq_{3}) G_{1,1}(\bk-\bq_{1}) G_{1,1}(\bk '-\bq_{2}) G_{1,1}(-\bk-\bq_{3}) G_{1,1}(-\bk'+\bq_{1}+\bq_{2}+\bq_{3}) \notag \\
+ & 4 \int d^{3}q_{1} \int d^{3}q_{2} B(\bq_{1},\bq_{2}) G_{1,2}(\bk+\bk' - \bq_{1})G_{1,1}(-\bk-\bq_{2})G_{1,1}(-\bk'+\bq_{1}+\bq_{2}) \notag \\
+ & \int d^{3}q_{1} \int d^{3}q_{2} B(\bq_{1},\bq_{2}) G_{1,2}(-\bq_{1})\biggl[G_{1,1}(\bk-\bq_{2})G_{1,1}(-\bk+\bq_{1}+\bq_{2}) + G_{1,1}(\bk'-\bq_{2})G_{1,1}(-\bk'+\bq_{1}+\bq_{2})\biggl] \notag \\
+ & 2 \int d^{3}q_{1} P(\bq_{1}) \biggl[G_{1,3}(\bk-\bq_{1})G_{1,1}(-\bk+\bq_{1}) + G_{1,3}(\bk'-\bq_{1})G_{1,1}(-\bk'+\bq_{1}) + |G_{1,2}(\bk-\bk'-\bq_{1})|^{2}\biggl] \notag \\
+ & \int d^{3}q_{1} P(\bq_{1}) |G_{1,2}(\bq_{1})|^{2} + (1+\alpha^{3})G_{1,4}(0) \biggl\},
\label{eq:appfinal}
\end{align}
\end{widetext}

where $P(\bq_{1})$, $B(\bq_{1},\bq_{2})$ and $T(\bq_{1},\bq_{2},\bq_{3})$ are the power spectrum, bispectrum and trispectrum respectively, $\alpha$ is the ratio of real to synthetic galaxies, and we have defined 
\begin{equation}
G_{\ell,m}(\bk) = \frac{\int d^{3}r \bar{n}^{\ell}(\br)w^{m}(\br)e^{i\bk\cdot\br}}{\biggl[ \int d^{3}r\bar{n}^{2}(\br)w^{2}(\br)\biggl]^{\frac{m}{2}}}.
\label{eq:appgterms}
\end{equation}
We have re-derived and shortened this expression compared to that given in \cite{Smith2016} by using the symmetry of the $\bk$-vectors and setting $\bk \rightarrow -\bk$ (and similarly for $\bk'$) in some of the individual terms.

\section{New approximation for the masked, binned covariance matrix} \label{sec:appcov2}
In this section we mathematically derive an approximation for the binned covariance matrix that would be measured from a set of masked mocks covering the full survey volume, i.e., Eq.~\ref{eq:appfinal}, in terms of only the binned covariance matrix measured from a set of cubic mocks $\bC^{sm}$ (the expression for which is given in Eq.~\ref{eq:covnowin}), and the analytically calculated Gaussian part of the convolved covariance matrix $\bC^{W}$ (the expression for which is given in Eq.~\ref{eq:Cov_cstPij}).

Starting with our assumption that the convolution with the window function is negligible for all terms in the covariance matrix except for the Gaussian part, we can write the first term in Eq.~\ref{eq:appfinal} as $\bC^{W}$. For the remaining terms we can perform a change of basis $\bq \rightarrow \bk-\bq$ and take the case that the $G$-terms are only non-zero for $\bq \approx 0$, i.e., that the window function only has support on large scales. Doing this for each of the higher-order trispectrum, bispectrum, power spectrum and constant terms results in
\begin{align}
\mathsf{C}(k_{i},k_{j}) & = \mathsf{C}^{W}(k_{i},k_{j}) + \bar{T}(k_{i},k_{j})G_{4,4}(0) \notag \\
& + \big[4\bar{B}(k_{i},k_{j})+\bar{B}(0,k_{j})+\bar{B}(k_{i}, 0)\big]G_{3,4}(0)  \notag \\
& + \big[\bar{P}(k_{i})+\bar{P}(k_{j})+\bar{P}(k_{i},k_{j})\big]G_{2,4}(0) \notag \\
& + \big[1+\alpha^{3}\big]G_{1,4}(0),
\label{eq:app2_1}
\end{align}
as the integrals over the $\bq$-vectors reduce to Dirac delta functions. For a constant number density, such as that in a simulation, the higher-order terms in Eq.~\ref{eq:app2_1} reduce exactly to those in Eq.~\ref{eq:covnowin}. However, even if this is not the case, we can substitute Eq.~\ref{eq:covnowin} into Eq.~\ref{eq:app2_1}. We choose to make this substitution about the trispectrum as this is the dominant higher-order term. The bispectrum, power spectrum and constant terms have a stronger dependence on the shot-noise and so for any modern large scale structure survey only become important on increasingly non linear scales. We can write the result of our substitution as
\begin{widetext}
\begin{equation}
\mathsf{C}(k_{i},k_{j}) = \mathsf{C}^{W}(k_{i},k_{j}) + \mathsf{C}^{res}(k_{i},k_{j}) + VG_{4,4}(0)
\biggl[\mathsf{C}^{sm}(k_{i},k_{j}) - \frac{2(2\pi)^{3}}{V_{k_{i}}V}\biggl(\bar{P}(k_{i})+\frac{1}{\bar{n}}\biggl)^{2}\delta^{D}(k_{i}-k_{j})\biggl],
\label{eq:app2_2}
\end{equation}  
where
\begin{align}
\mathsf{C}^{res}(k_{i},k_{j}) & =  \biggl[1+\alpha^{3}\biggl]\biggl(G_{1,4}(0) - \frac{G_{4,4}(0)}{\bar{n}^{3}}\biggl) + \biggl[\bar{P}(k_{i})+\bar{P}(k_{j})+\bar{P}(k_{i},k_{j})\biggl]\biggl(G_{2,4}(0) - \frac{G_{4,4}(0)}{\bar{n}^{2}}\biggl) \notag \\
& + \biggl[4\bar{B}(k_{i},k_{j})+\bar{B}(0,k_{j})+\bar{B}(k_{i}, 0)\biggl]\biggl(G_{3,4}(0)-\frac{G_{4,4}(0)}{\bar{n}}\biggl).
\end{align}
\end{widetext}
We have dubbed $\bC^{res}$, the \textit{residual} term. For our work we make the approximation that this is zero (which is actually true if the number density is constant), and find very good agreement in the case of the MGS mocks. One could instead measure the bispectrum and power spectrum terms from the set of small volume mock catalogues that are used to estimate the small scale covariance matrix (indeed this could also be done for the trispectrum, bypassing the substitution altogether) and computing the necessary scaling factors. However, as our current method is already accurate enough and computing the higher order clustering of every mock can be computationally demanding we do not attempt this here.

\end{document}